\documentstyle[11pt]{article}

\def\version{  version 9.2  --  28-2-94  -- }
\catcode`\@=11
\newif\if@fewtab\@fewtabtrue
{\count255=\time\divide\count255 by 60
\xdef\hourmin{\number\count255}
\multiply\count255 by-60\advance\count255 by\time
\xdef\hourmin{\hourmin:\ifnum\count255<10 0\fi\the\count255}}
\def\ps@draft{\let\@mkboth\@gobbletwo
    \def\@oddhead{}
    \def\@oddfoot {\hbox to 7 cm{\tiny \version \draftdate
       \hfil}\hskip -7cm\hfil\rm\thepage \hfil}
    \def\@evenhead{}\let\@evenfoot\@oddfoot}

\def\yes{yes }
 \read-1 to\yesno
\ifx\yesno\yes 
\font\tendl=msym10  scaled \magstep1
\font\sevendl=msym7 scaled \magstep1
\font\fivedl=msym5 scaled \magstep1
\newfam\dlfam \def\dl{\fam\dlfam\tendl}
\textfont\dlfam=\tendl \scriptfont\dlfam=\sevendl
\scriptscriptfont\dlfam=\fivedl
\else \let\dl=\bf
\fi

\def\ifundefined#1{\expandafter\ifx\csname#1\endcsname\relax}
\makeatletter \ifundefined{new@mathgroup} {} \else
    \new@mathgroup\sffam
    \new@internalmathalphabet\mathsf\sffam{cmss}{m}{n}
    \def\psf{\fontfamily\sfdefault \fontseries\default@series
        \fontshape\default@shape\selectfont\mathsf}
\fi

\def\citen#1{\if@filesw \immediate\write \@auxout {\string\citation{#1}}\fi%
\@tempcntb\m@ne \let\@h@ld\relax \def\@citea{}%
\@for \@citeb:=#1\do {\@ifundefined {b@\@citeb}%
    {\@h@ld\@citea\@tempcntb\m@ne{\bf ?}%
    \@warning {Citation `\@citeb ' on page \thepage \space undefined}}%
    {\@tempcnta\@tempcntb \advance\@tempcnta\@ne
    \setbox\z@\hbox\bgroup\ifcat0\csname b@\@citeb \endcsname \relax
    \egroup \@tempcntb\number\csname b@\@citeb \endcsname \relax
    \else \egroup \@tempcntb\m@ne \fi \ifnum\@tempcnta=\@tempcntb
    \ifx\@h@ld\relax \edef \@h@ld{\@citea\csname b@\@citeb\endcsname}%
    \else \edef\@h@ld{\hbox{--}\penalty\@highpenalty
    \csname b@\@citeb\endcsname}\fi
    \else \@h@ld\@citea\csname b@\@citeb \endcsname \let\@h@ld\relax \fi}%
\def\@citea{,\penalty\@highpenalty\hskip.13em plus.13em minus.13em}}\@h@ld}
\def\@citex[#1]#2{\@cite{\citen{#2}}{#1}}%
\def\@cite#1#2{\leavevmode\unskip\ifnum\lastpenalty=\z@\penalty\@highpenalty\fi%
  \ [{\multiply\@highpenalty 3 #1%
  \if@tempswa,\penalty\@highpenalty\ #2\fi}]}   %
\makeatother 

\def\draftdate{\number\month/\number\day/\number\year\ \ \ \hourmin }

\catcode`\@=12

\def\alg           {algebra}
\newcommand{\andauthoretc}[5]{\\[7 mm]{} \cli{\sc #1}
                   \\[2 mm] \cli{#2}\\[.5 mm] \cli{#3}\\[.5 mm]
                   \cli{#4}\\[.5 mm] \cli{#5}}
\def\Ao            {\mbox{$A_1^{(1)}$}}

\def\apo           {\mbox{$S$}}
\long\def\authoretc#1#2#3#4#5{\cli{\sc #1}\\[2 mm]{} \cli{#2}\\[.5 mm]
                   \cli{#3}\\[.5 mm] \cli{#4}\\[.5 mm] \cli{#5}}
\def\be            {\begin{equation}}
\def\bfe           {{\bf1}}
\let\Bi=\bibitem
\def\bmeG          {\bar\zeta}
\def\bmega         {\bar\omega}
\def\BOH           {\mbox{${\cal B}({\cal H}_\pi)$}}
\def\bp            {{\bar p}}
\def\bq            {{\bar q}}
\def\bzeta         {\bar\zeta}
\def\cdo           {\cdot}
\def\cft           {conformal field theory}
\def\cfts          {conformal field theories}
\def\CG            {Clebsch\hy Gor\-dan }
\def\CGC           {Clebsch\hy Gor\-dan coefficient}
\newcommand{\cgc}[7]{\mbox{{\Large[}$\begin{array}{ccc} {}\\[-1.55em]
     \!\!\scs#1&\!\!\!\! \scs#2&\!\!\scs#3\!\! \\[-.32em] \!\!\scs#4
     &\!\!\!\!\scs#5&\!\!\scs#6\!\! \end{array}{\mbox{\Large]}}^{}_{#7}$\,}}

\newcommand{\cgcs}[7]{\mbox{{\Large[}$\begin{array}{ccc} {}\\[-1.55em]
     \!\!\scs#1&\!\!\!\! \scs#2&\!\!\scs#3\!\! \\[-1.04mm] \!\!\scs#4
     &\!\!\!\!\scs#5&\!\!\scs#6\!\! \end{array}{\mbox{\Large]}}^*_{#7}$\,}}

\let\cli=\centerline
\def\Co            {{\dl C}}
\def\coa           {\mbox{$\varphi$}}
\def\coc           {\mbox{$\cal R$}}

\def\complex       {{\dl C}}
\def\cop           {\mbox{$\Delta$}}
\def\copp          {\mbox{$\Delta'$}}
\newcommand{\copmatab}[1] {\begin{array}{ccc} &\,\,\ 0&1\,\,\ \\[2 mm]
        \begin{array}{c} 0\\1 \end{array} &
        \left( \begin{array}{c} 0\\1 \end{array} \right. &
        \left. \begin{array}{c} #1 \end{array} \right) \end{array} }
\newcommand{\copmatabb}[2] {\begin{array}{cccc} &\,\,\ 0&1_1&1_2\,\,\ \\[2 mm]
        \begin{array}{c}0\\1_1\\1_2\end{array} &
        \left( \begin{array}{c} 0\\1_1\\1_2 \end{array} \right. &
               \begin{array}{c} #1 \end{array} &
        \left. \begin{array}{c} #2 \end{array} \right) \end{array} }
\newcommand{\copmatabc}[2] {\begin{array}{cccc} &\,\,\ 0&1&2\,\,\ \\[2 mm]
        \begin{array}{c}0\\1\\2\end{array} &
        \left( \begin{array}{c} 0\\1\\2 \end{array} \right. &
               \begin{array}{c} #1 \end{array} &
        \left. \begin{array}{c} #2 \end{array} \right) \end{array} }
\newcommand{\copmatabbccc}[5] {\begin{array}{ccccccc} &\,\,\ 0&1_1&1_2&2_1
        &2_2&2_3\,\,\ \\[2 mm] \begin{array}{c}0\\1_1\\1_2\\2_1\\2_2\\2_3
        \end{array} & \left( \begin{array}{c} 0\\1_1\\1_2\\2_1\\2_2\\2_3
        \end{array} \right. & \begin{array}{c} #1 \end{array} &
        \begin{array}{c} #2 \end{array} & \begin{array}{c} #3 \end{array} &
               \begin{array}{c} #4 \end{array} &
        \left. \begin{array}{c} #5 \end{array} \right) \end{array} }
\newcommand{\copmatabbccd}[5] {\begin{array}{ccccccc} &\,\,\ 0&1_1&1_2&2_1
        &2_2&3 \,\,\ \\[2 mm] \begin{array}{c}0\\1_1\\1_2\\2_1\\2_2\\3
        \end{array} & \left( \begin{array}{c} 0\\1_1\\1_2\\2_1\\2_2\\3
        \end{array} \right. & \begin{array}{c} #1 \end{array} &
        \begin{array}{c} #2 \end{array} & \begin{array}{c} #3 \end{array} &
               \begin{array}{c} #4 \end{array} &
        \left. \begin{array}{c} #5 \end{array} \right) \end{array} }
\newcommand{\copmatabcc}[3] {\begin{array}{ccccc} &\,\,\ 0&1&2_1&2_2\,\,\
        \\[2 mm] \begin{array}{c}0\\1\\2_1\\2_2\end{array} &
        \left( \begin{array}{c} 0\\1\\2_1\\2_2 \end{array} \right. &
        \begin{array}{c} #1 \end{array} & \begin{array}{c} #2 \end{array} &
        \left. \begin{array}{c} #3 \end{array} \right) \end{array} }
\def\cou           {\mbox{$\epsilon$}}
\newcommand{\dbx}[1] {\dashbox{1.8}(10,10){$#1$}}
\def\degen         {{\small\rm degenerate}}
\newcommand{\del}[1] {}

\def\DHR           {Dop\-li\-cher\hy Haag\hy Ro\-berts }
\let\dstyle=\displaystyle
\newcommand{\e}[3] {\mbox{$e^{#2, #3}_{#1}$}}
\def\ee            {\end{equation}}

\newcommand{\epij}[1]{\mbox{$e^{i_{#1},j_{#1}}_{p_{#1}}$}}

\newcommand{\erf}[1]{(\ref{#1})}
\newcommand{\etij}[1]{\mbox{$e^{i_{#1},j_{#1}}_{t_{#1}}$}}
\newcommand{\F}[8] {\mbox{$F^{(#1)_{\scriptstyle#8}}_{ {#2} #3 {#4},
                   {#5} #6 {#7}}$}}
\newcommand{\Fa}[8]{\mbox{$F^{(#1)_{\scriptstyle#8}}_{ {#4} #3 {#2},
                   {#7} #6 {#5}}$}}
\def\FA            {\mbox{$\cal F$}}
\newcommand{\Fb}[8]{\mbox{$\overline F^{\,(#1)_{\scriptstyle#8}}_{{#2}#3{#4},
                   {#5}#6{#7}}$}}
\def\Fbpqr         {\Fb{pqr}\alpha u\beta\gamma v\delta t}
\newcommand{\fbx}[1] {\framebox(10,10){$#1$}}
\newcommand{\FF}[6]{\mbox{$F^{(#1#2#3)_{\scriptstyle#6}}_{#4,#5}$}}
\newcommand{\FFb}[6]{\mbox{$\overline
F^{\,(#1#2#3)_{\scriptstyle#6}}_{#4,#5}$}}
\newcommand{\Fh}[8] {\mbox{$\hat F^{(#1)_{\scriptstyle#8}}_{ {#2} #3 {#4},
                   {#5} #6 {#7}}$}}
\newcommand{\Fha}[8]{\mbox{$\hat F^{(#1)_{\scriptstyle#8}}_{ {#4} #3 {#2},
                   {#7} #6 {#5}}$}}
\newcommand{\Fhb}[8]{\mbox{$\overline{\hat F}\!{}^{\,(#1)_{\scriptstyle#8}}_
                   {{#2}#3{#4}, {#5}#6{#7}}$}}
\def\Fhpqr         {\Fh{pqr}\alpha u\beta\gamma v\delta t}
\def\findim        {fini\-te-dimen\-si\-onal}
\newcommand{\fline}[1]{\vfill\noindent ------------------\\[1 mm]}
\def\Fpqr          {\F{pqr}\alpha u\beta\gamma v\delta t}
\def\furu          {fusion rule}
\def\futnot#1      {}
\def\futnote#1     {\footnote{~#1}\ }
\def\h             {\mbox{$H$}}
\let\H=\h
\def\heq           {\,\hat=\,}
\def\hh            {\hat H}
\def\hhh           {|\hat H|}
\def\HS            {\mbox{$\cal H$}}
\newcommand{\hsp}[1] {\mbox{\hspace{#1 em}}}
\def\hy            {$\mbox{-\hspace{-.66 mm}-}$}
\newcommand{\hz}[3]{\mbox{$\h[(#1_{#2})^{}_{#3}]$}}
\newcommand{\hzz}[6]{\mbox{$\h[(#1_{#2})^{}_{#3}\!\oplus\!(#4_{#5})^{}_{#6}]$}}
\def\id            {{\sl id}}
\def\ii            {{\rm i}}

\def\interpr       {\multicolumn{1}{c} {\small interpretation}}
\def\irrep         {irreducible representation}
\def\kma           {Kac\hy Moody algebra}
\long\def\labl#1   {\label{#1}\ee} 
\def\largeformat   {\addtolength{\oddsidemargin}{-2.02 cm}
                   \addtolength{\topmargin}{-2.6 cm}
                   \setlength{\textwidth}{15.8 cm}}

\def\Llb           {\mbox{\Large(}}
\def\Lrb           {\mbox{\Large)}}
\def\LY            {Lee\hy$\!$Yang }

\def\MS            {{\tt S}}
\def\MT            {{\tt T}}

\def\npqr          {\mbox{$N_{pq}^{\ \,r}$}}
\def\nrst          {\mbox{$N_{rs}^{\ \,t}$}}
\def\OA            {\mbox{$\cal A$}}
\def\omeG          {\zeta}
\def\one           {\mbox{\small $1\!\!$}1}
\def\onedim        {one-di\-men\-si\-o\-nal}

\def\oti           {\otimes}
\def\otim          {\otimes}
\def\otiM          {\!\otimes\!}
\def\PA            {{\bf p}}
\def\PL            {{\bf p}} 
\def\PR            {{\bf p}} 
\def\qft           {quantum field theory}
\def\qfts          {quantum field theories}
\def\qg            {quantum group}
\def\qs            {quantum symmetry}
\def\qss           {quantum symmetries}
\long\def\query#1{\hskip 0pt{\vadjust{\everypar={}\small\vtop to 0pt{\hbox{}%
     \vskip -13pt\rlap{\hbox to 46.3pc{\hfil{\vtop{\hsize=8pc\tolerance=6000%
     \hfuzz=.5pc\rightskip=0pt plus 3em\noindent#1}}}}\vss}}}}%
\newcommand{\R}[5] {\mbox{$R^{(#1 #2)_{\scriptstyle#3}}_{#4, #5}$}}
\def\reals         {{\dl R}}
\def\rep           {representation}
\def\Rep           {Representation}
\def\rha           {rational Hopf algebra}
\let\RHA=\rha
\newcommand{\RR}[3]{\mbox{$R^{(#1 #2)_{\scriptstyle#3}}_{}$}}
\newcommand{\RRb}[3]{\mbox{$\overline R^{\,(#1 #2)_{\scriptstyle#3}}_{}$}}
\def\rqft          {rational quantum field theory}
\def\rqfts         {rational quantum field theories}
\let\scs=\scriptstyle
\newcommand{\sect}[1] {\section{#1}\setcounter{equation}{0}}

\def\sigMa         {\sigma}
\def\sigmA         {\sigma}
\def\sigmASQ       {\parallel\!H\!\parallel}
\def\sigmASQI      {\parallel\!H\!\parallel^{-1}}
\def\sqf           {\frac{1+\sqrt5}2}
\def\stop          {\mbox{$\epsilon$}}
\def\stopt         {\mbox{$\epsilon$}}

\newcommand{\sumh}[1]{\sum_{#1\in\hat H}}
\newcommand{\sumi}[2]{\sum_{#1=1}^{n_{#2}}}
\newcommand{\sumn}[4]{\sum_{#1=1}^{N_{#2 #3}^{\ \,#4}}}

\def\tfts          {topological field theories}
\def\tims          {\times}
\newcommand{\titleetC}[6] { {\tt #1}\mbox{}\\[-13 mm]
     \rightline{{\sf NIKHEF-H/94-05}} \\ \rightline{{\sf KL-TH-94/4}} \\
     \rightline{{\sf hep-th/9402153}}\\
     \rightline{{\sf #2}}\\[11 mm]%
     \cli{\bf\Large {#3}}\\[2.3 mm] \cli{\bf\Large {#4}}\\[11 mm]{}%
     {#5} \\[10 mm]{}%
     \begin{quote}{\bf Abstract.}\ \,{#6}\end{quote} \newpage}%
\def\twodim        {two-di\-men\-si\-o\-nal}
\def\typeversion {\typeout{}\typeout{ ---- \version ---- }\typeout{}}
\def\U             {{V}}
\let\UN=\bfe

\newcommand{\wz}[3]{\mbox{$(#1_{#2})^{}_{#3}$}}
\newcommand{\wZ}[3]{\mbox{$(#1_{#2}^{(1)})^{}_{#3}$}}

\def\wzwm          {WZW model}

\def\wzwts         {WZW theories}
\def\zet           {{\dl Z}}
\def\zeT           {\omega}
\let\zetab=\zeta

\def\zets          {{\scriptstyle\dl Z}}
\def\zz            {z}
\def\bP  {\begin{picture}}
\def\bPo {\begin{picture}(0,0)}
\def\eP  {\end{picture}}

\newcommand{\mpliv}[9] {\multiput(#1,#2)(#3,#4){#5}{\line(#6,#7){#8}}
                       \multiput(#1,#2)(#3,#4){#5}{\vector(#6,#7){#9}}}

\newcommand{\pulin}[5] {\put(#1,#2){\line(#3,#4){#5}}}
\newcommand{\puliv}[6] {\put(#1,#2){\line(#3,#4){#5}}
                       \put(#1,#2){\vector(#3,#4){#6}}}
\newcommand{\puova}[5] {\put(#1,#2){\oval(#3,#4)[#5]}}

\newcommand{\putsc}[3] {\put(#1,#2){{\scriptsize $#3$}}}
\newcommand{\putsm}[3] {\put(#1,#2){{\small $#3$}}}

\newcommand{\puvec}[5] {\put(#1,#2){\vector(#3,#4){#5}}}
\newcommand{\tetrA}[9]
  {\bP(80,63)  \puliv{25}{25}01{30}{18}
  \puliv{-5}{10}01{30}{27.4} \puliv{55}{40}0{-1}{30}{6}
  \puova{25}{40}{60}{30}t \puova{25}{10}{60}{30}b
  \pulin{25}{25}{5}{-3}{29.55} \pulin{25}{25}{-5}{-3}{29.55}
  \puvec{36.2}{18.4}3{-2}5 \puvec{7.3}{14.3}325 \puvec{27.4}{-5}{-1}05
  \putsc{-10.7}{5.5}{#9} \putsc{22.5}{18.8}{#7} \putsc{23.1}{57.9}{#8}
  \putsm{-13.9}{35}{#1} \putsm{59.3}{35}{#4} \putsm{22.9}{-15.8}{#6}
  \putsm{9}{20}{#5} \putsm{38}{19.8}{#3} \putsm{28}{40}{#2} \eP}

\newcommand{\tetrab}[9]
  {\bP(96,63)  \puliv{25}{-5}01{30}{18} \mpliv{-5}{10}{60}0201{30}{9.4}
  \puova{25}{40}{60}{30}t \puova{25}{10}{60}{30}b
  \pulin{25}{25}53{29.55} \pulin{25}{25}{-5}3{29.55}
  \puvec{13.9}{31.4}{-3}25 \puvec{42.2}{35.2}{-3}{-2}5 \puvec{26.8}{54.7}{-1}05
  \putsc{3.8}{6.9}{\dbx{#8}}\putsc{36.3}{6.9}{\dbx{#7}} \putsc{19.8}{37.4}{#9}
  \putsm{-13.4}{13.5}{#6} \putsm{59.2}{13.5}{#1} \putsm{22.8}{58.1}{#5}
  \putsm{6.9}{26.5}{#3} \putsm{38.7}{26.5}{#2} \putsm{27.5}{10.8}{#4} \eP}
\newcommand{\tetraf}[9]
  {\bP(96,63)  \puliv{25}{25}01{30}{18}
  \puliv{-5}{10}01{30}{27.4} \puliv{55}{40}0{-1}{30}{6}
  \puova{25}{40}{60}{30}t \puova{25}{10}{60}{30}b
  \pulin{25}{25}5{-3}{29.55} \pulin{25}{25}{-5}{-3}{29.55}
  \puvec{36.2}{18.4}3{-2}5 \puvec{13.9}{18.4}{-3}{-2}5 \puvec{23.2}{-5}105
  \putsc{3.8}{34.3}{\fbx{#7}}\putsc{36.3}{34.3}{\fbx{#8}} \putsc{19.8}{2.6}{#9}
  \putsm{-13.4}{35}{#2} \putsm{59.4}{35}{#3} \putsm{22.8}{-15.6}{#5}
  \putsm{7.4}{20.3}{#1} \putsm{38.7}{20.3}{#6} \putsm{27.0}{42.8}{#4} \eP}

\largeformat
\ifx\yesno\yes 
\else  \fi

\begin{document} 
\titleetC {} {February 1994}
 {TOWARDS A CLASSIFICATION} {OF RATIONAL HOPF ALGEBRAS}
 {\authoretc{\ J\"urgen Fuchs\ $^1$}{NIKHEF-H}{Kruislaan 409}
    {NL -- 1098 SJ~~Amsterdam}{The Netherlands}
  \andauthoretc{\ Alexander Ganchev $^2$}{Fachbereich Physik
    -- Theoretische Physik} {Erwin Schr\"odinger-Stra\ss e 46/55}
  {D -- 67653~~Kaiserslautern} {Germany}
  \andauthoretc{\ Peter Vecserny\'es $^3$}{Central Research Institute
    for Physics} {P.O.\ Box 49}{H -- 1525~~Budapest \,114}{Hungary} }
 {Rational Hopf algebras, i.e.\ certain quasitriangular weak quasi-Hopf
\mbox{$^*$-algebras,} are expected to describe the quantum symmetry of
rational field theories. In this paper
methods are developped which allow for a classification of all
rational Hopf algebras that are compatible with some prescribed set of
fusion rules. The algebras are parametrized by the solutions of
the square, pentagon and hexagon identities.
As examples, we classify all solutions for fusion rules with not
more than three sectors, as well as for the level three affine $A_1^{(1)}$
fusion rules. We also establish several general properties of rational
Hopf algebras and present a graphical description of the coassociator in
terms of labelled tetrahedra. The latter construction allows to make
contact with \cft\ fusing matrices and with invariants of three-manifolds
and topological lattice field theory.
 {}\fline{} {\small $^1$~~Heisenberg fellow\\[.2 mm]
 $^2$~~Humboldt fellow; on leave from INRNE, Sofia, Bulgaria \\[.2 mm]
 $^3$~~\,partially supported by Fulbright Foundation and the Hungarian
       Scientific Research Fund \\ \hsp{.75} (OTKA 1815) }}


\sect{Introduction}

Rational Hopf algebras have been proposed as candidates for the
description of the superselection symmetries of rational quantum field
theories.  In this paper we present further evidence that these \alg s can
indeed play such a role. To do so, we describe as examples the \rha s that
reproduce the fusion rules, the conformal weights,
the quantum dimensions and the representation of the modular group for a
variety of (chiral) \cfts, among them e.g.\ the Ising model and the
\wz G21, \wz F41, \wz Br1 and \wz A13 \wzwts.

In particular, we obtain a complete classification of all \rha s \h\
for which the number $\hhh$ of simple ideals of \h\ does not exceed three.
More generally, we describe an algorithm by which one obtains, by
starting from an arbitrary candidate set of \furu s, a complete
classification of all \rha s that are compatible with these given \furu s.
Furthermore, we derive several general results about \rha s, among them
the commutativity and associativity of the \furu s
and simple general formul\ae\ for the statistics operators and for the
monodromy matrix. We also introduce a graphical representation of
the coassociator in
terms of tetrahedra which clarifies the relation with \cft\ and with
triangulations of three-manifolds and topological field theory.

\vskip 2mm
The aim of the \DHR programme \cite{DHR} is to explore the symmetries and the
statistics properties of quantum field theoretical models
of point particles with localized charges by using `observable' data alone.
That programme has been
completed \cite{DR} in $D\geq3$ dimensions of Minkowski space-time.
The superselection sectors can be characterized by representations of
compact groups, and they must satisfy Bose or Fermi statistics.
As a consequence, the classification of compact groups leads to a
partial classification of the possible \qfts\ in Minkowski space-time
of dimension $D\geq3$. Owing to the simple braiding structure
compatible with compact groups, this classification is only rather poor.
In contrast, in two dimensions a much richer structure,
braid group statistics, is allowed. Therefore when restricting attention
to field theories with `fully braided' superselection sectors one
expects that the classification of the underlying `quantum symmetries'
which replace the compact groups in low number of dimensions leads to
a considerably finer classification of the allowed field theories.
Unfortunately, the quantum symmetry which is dual to the superselection
structure of a low-dimensional \qft\ is not yet known in general;
however, several proposals for quantum symmetries have been made
\cite{masc,rehr,haya,kerl,V}.
In particular, for {\em \rqfts}, certain bi-algebras known as
{\em \rha s\/} have been suggested (\hsp{-.37}\cite V; compare also
\cite{S,SV}) as the relevant \qss; here by a \rqft\ we mean a
theory with a finite number of superselection sectors, among which
there is a single degenerate sector (the vacuum sector), i.e.\ a
single sector that has trivial monodromy with all of the sectors.
In the presence of such a symmetry the
superselection sectors carry a unitary representation of the modular group
$\Gamma=SL(2,\zet)$ (even in the absence of conformal invariance), and the
representation matrices for the generators of $\Gamma$ are given
in terms of the monodromy matrix and the statistics phases of the model,
respectively.

As the multiplicities of irreducible \DHR$\!$-\rep s of the observable algebra
are finite, the \rep s of the quantum symmetry \h\ that correspond
to the sectors are \findim. In the case of a \rqft, this implies
that \h\ must be \findim\ as well. In the reconstruction of a particular
model we would like to have a field algebra \FA\ on which the symmetry
algebra \h\ acts in such a way that the observable
algebra \OA\ is the \h-invariant subalgebra of \FA. The inequivalent
representations $\pi_p$ of \OA\ (with certain multiplicities)
arise as sub-representations of the representation $\pi$ of \FA\ on
the Hilbert space $\HS_\pi$;
they are in one to one correspondence with equivalence classes
$[\rho_p]$ of endomorphisms $\rho_p$ of \OA\ through the vacuum
representation $\pi_0$ of \OA, according to $\pi_p\simeq\pi_0\circ\rho_p$.
In order to have a faithful realization $U$ of \h\ in \BOH\ such that
$U(\h)$ is the commutant of $\pi(\OA)$,
$U(\h)$ must be a von Neumann algebra; this implies that \h\ should be a
\findim\ semisimple algebra, and hence a finite direct sum of full
matrix algebras,
  \be  \h=\bigoplus_{p\in\hh}M_{n_p}^{}   \labl{h=+}
with $\hh$ a finite index set.
In particular, $U(\h)$ and $\pi(\OA)^-=U(\h)'$ have a common center
so that the minimal central projectors of \h\ (which lead to the
inequivalent representations of \h) are in one to one correspondence with the
inequivalent representations of the observables.

Various properties of the \rep s $\pi_p$ of \OA\ play an essential role in
the \DHR programme. For compatibility, i.e.\ for categorical equivalence,
analogous properties must hold for the \rep s of the quantum symmetry \h.
The relevant requirements can be summarized as follows.
There exist \findim\ irreducible representations; a product of
\rep s is defined (which is unique up to unitary equivalence), and the
product of \irrep s is completely reducible into \irrep s.
There exists a one-dimensional vacuum representation, which plays
the role of a unit when taking products of representations.
There is the notion of contragredient (or conjugate) representation
(unique up to unitary
equivalence); the product of a representation with its contragredient
contains the trivial representation, with multiplicity one in the case of
irreducible representations. The product of representations is commutative
and associative up to unitary equivalence.
The commutativity and associativity isomorphisms obey a set of
compatibility relations, which in a category theoretic setting mean that
the representations of \h\ and their intertwiners constitute the objects
and arrows, respectively, of a braided monoidal rigid $C^*$-category.
Finally, the irreducible representations of \h\ induce
a $\hhh$-dimensional unitary representation of the modular group $\Gamma$.

In order that the coproduct of \h\ be compatible with the fusion rules
  \be  [\rho_p\circ\rho_q]=\sum_r\npqr\,[\rho_r] \,, \labl{furu}
the dimensionalities $n_p$ of the ideals $M_{n_p}$ that appear in
the sum \erf{h=+} must satisfy the requirements that $n_0=1$ and $n_\bp^{}
=n_p$ (here $p\mapsto \bar p$ is the involution describing the `charge
conjugation' among isomorphic minimal ideals 
 in \H\ that corresponds to the conjugation of \h-\rep s),
as well as the inequality
  \be  n_p n_q\geq\sumh r\npqr\,n_r \,,  \labl{npqr}
which is needed to `accomodate without overlapping' the matrix
algebra $\bigoplus_r(\bigoplus_{\alpha=1}^{N_{pq}^{\ \;r}}M_{n_r}^
{(\alpha)})$ into $M_{n_p}\oti M_{n_q}$.
As it turns out (see below), the structure of \rha s does not put
any further restrictions on the integers $n_p$. Some implications of this
observation will be discussed in section \ref{secnp}.

In order for \h\ to reproduce not only the fusion rules \erf{furu},
but also the associated braid group representations,
statistics parameters and monodromy matrix, one
studies algebra embeddings $\nu$ of the form
  \be  \nu\colon\ \h\to M_n(\h) \,, \labl{amp}
where the elements of $M_n(\h)$ are $n\times n$ matrices with entries in $\h$.
Such amplifying monomorphisms or, shortly, {\em amplimorphisms\/} $\nu$ of \H\
mimic the endomorphisms $\rho$ of the observable algebra $\OA$; in particular,
they have trivial kernel, but they are automorphisms only
in the particular case of $n=1$. The composition of amplimorphisms
leads to an associative product $\tims$. The braiding properties
of $\tims$ are described by the {\em statistics operator\/} which
has similar properties as the statistics operator in algebraic
field theory. In particular, the statistics operators lead in a
natural manner to statistics parameters, coloured braid
group representations and to the monodromy matrix.

Amplimorphisms $\nu_r$ of a \RHA\ \H\ are naturally constructed
with the help of a coproduct \cop\ of \h, according to
  \be  \h\ni a\mapsto (\UN\otimes e_r)\cdot\cop(a)
  =\sumi{i,j}r \nu_r^{ij}(a)\otim\e rij\in\h\oti\h\,,\qquad
   \nu_r(a) \equiv [\nu_r^{ij}(a)] \in M_{n_r}(\H) \,.  \labl{emb}
Here by \e rij, $r\in\hh,\, i,j=1,\ldots ,n_r$, we denote the matrix units
which provide a linear basis of \h;
$e_r\in H$ denotes the central projector corresponding to a simple summand
$M_{n_r}$ of dimension $n_r$ of $H$, and $\UN=\sumh r e_r$ is the unit
element of \H. If an embedding of the form \erf{emb} is unit preserving, it
gives rise to an index $n_r^2$ \cite J,
while the `minimal' embedding $\H\mapsto \H\otim \e{}11$,
where $\e{}11$ is a matrix unit, yields index one. To reproduce a
non-integral statistical dimension $d_r$ of a sector $[\rho_r]$ (which is equal
to the square root of the corresponding index $[\OA : \rho_r(\OA)]$ of
the embedding of $\rho_r(\OA)$ into \OA, one is
forced to use a unit non-preserving coproduct in \H\ that leads to an
intermediate statistical dimension $1<d_r<n_r$.

A major step in the investigation of \qs\ will be the classification of the
\alg s \h\ that lead to the properties listed above. Such a classification is
highly non-trivial, and at present still beyond reach. As a first step
of the classification programme, we start in this paper from some fixed
system of fusion rules and determine the set of \rha s that is compatible
with them. Among the solutions, one expects in particular \rha s
corresponding to such (chiral halves of) unitary rational \cfts\ that
share the relevant fusion rules.

The \furu\ \alg s obtainable via
rational Hopf algebras are a subset of the so-called \cite F
{\em modular\/} fusion rule algebras. As a consequence, in the
classification we need to consider only such coproducts that correspond
to modular fusion rules; for other coproducts the full set of
conditions on \rha s will not have any solution. Modular fusion rules
have been completely classified for the case where the number $\hhh$ of sectors
is smaller or equal to three; we will describe all \rha s that correspond
to such fusion rules.

\vskip 2mm
The rest of this paper is organized as follows.
In section 2, following \cite V, we recall briefly the definition of
rational Hopf algebras and describe their gauge freedoms, the category of
amplifying morphisms of rational Hopf algebras, the standard left
inverse of an amplimorphism, and the monodromy matrix. In
section 3 we start by describing the coproduct \cop\ in terms of
\CGC s, and consequently express the cocommutator \coc\
and the coassociator \coa\ through linear combinations
of products of \CGC s with numerical coefficents
$\R rst\alpha\beta$ and \Fpqr, respectively. These numerical matrices
$R$ and $F$ are independent of the dimensionalities $n_p$
of the simple summands $M_{n_p}$ of \H. We then proceed to rewrite the
pentagon and hexagon equations, as well as the
statistics parameters and the monodromy matrix
in terms of the matrices $R$ and $F$, and hence in $n_p$-independent form.

A graphical (path) notation for various quantities of interest is
introduced in section 4; for example, the \CG operator,
or equivalently the coproduct, is
described by a trivalent vertex. The associator then corresponds to
certain labelled tetrahedra; working in the dual graphical description,
where a trivalent vertex is replaced by a triangle, the coassociator is
again expressed by tetrahedra, and the pentagon equation
can be interpreted as describing a move in triangulations of
three-manifolds. Motivated by this observation, we then show that the $F$
coefficients satisfy also an equation corresponding to a subdivision
move of the triangulation. As a consequence, we can associate with every
rational Hopf algebra an invariant of three-manifolds.

A complete classification of rational Hopf algebras with $|\hh|\le 3$
simple summands is obtained in section 5. In section 6 we classify the
4-sector rational Hopf algebras whose character ring equals the
fusion ring of the level 3 \Ao\ WZW model.
This fusion ring is the tensor product of two 2-element fusion rings,
and it turns out that this tensor product structure extends to the
rational Hopf algebras. The final section 7 contains a brief discussion
of our results and of some issues which deserve further investigation.

\sect{Rational Hopf algebras}

In this section we collect the relevant properties of \rha s,
their \rep s and the associated amplimorphisms; for more details, see
\cite{V,S,SV}.

In short, a \rha\ is a quasitriangular weak quasi-Hopf $^*$-algebra with
certain additional properties. More specifically,
the defining axioms of a \rha\ \h\ are the following:

\begin{itemize}
\item \H\ is an associative \findim\ semisimple ${}^*$-algebra
with unit, and hence a direct sum \erf{h=+} of full matrix \alg s.
\h\ has a \onedim\ \rep\ $\cou\colon\ \H\to\complex$,
called the co-unit, which is a unit preserving ${}^*$-homomorphism,
and is endowed with a coproduct $\cop\colon\ \H\to \H\oti\H$,
which is a ${}^*$-monomorphism. \\(The $^*$-operation on $\h\oti\h$ is
defined by $(a\otimes b)^*=a^*\otimes b^*$.)

\item There are unitary elements $\rho,\lambda\in\H$ such that
  \be  (\cou\oti\id)\circ\cop(a)=\rho a \rho^*, \quad\qquad
  (\id\oti\cou)\circ\cop(a)=\lambda a\lambda^* \labl{rar}
for all $a\in\h$.

\item \h\ is endowed with a linear ${}^*$-antiautomorphism
  $\apo\!:\ \h\to\h$, called the antipode.
There exist non-zero elements $l,r\in\H$ such that
  \be  a^{(1)}\cdot l\cdot \apo(a^{(2)})=l\cdot\cou(a),\quad\qquad
  \apo(a^{(1)})\cdot r\cdot a^{(2)} =\cou(a)\cdot r \labl{alr}
for all $a\in\h$. Here for brevity we use Sweedler type notation, i.e.\
write $\cop(a)=a^{(1)}\otim a^{(2)}$ in place of
$\cop(a)=\sum_i^{} a_i^{(1)}\otim a_i^{(2)}$.

\item The coproduct is quasi-cocommutative. Thus there is an element
$\coc\in\H\oti\H$ such that
  \begin{eqnarray} && \copp(a)\cdot\coc= \coc\cdot\cop(a)\ \qquad \mbox{
  for all }a\in\h, \label{coc1} \\[1.4 mm]&&
  \copp(\bfe)\cdot\coc= \coc =\coc\cdot\cop(\bfe), \label{coc2} \\[1.4 mm]&&
  \coc\cdot\coc^*=\copp(\bfe),\qquad \coc^*\cdot\coc=\cop(\bfe). \label{coc3}
  \end{eqnarray}
Here $\copp\equiv\tau\circ\cop$ denotes the opposite coproduct.

\item The coproduct is quasi-coassociative. Thus there is an element
$\coa\in\H\oti \H\oti\H$ such that
  \begin{eqnarray} && (\cop\oti\id)\circ\cop(a)\cdot\coa =
  \coa\cdot(\id\oti\cop)\circ\cop(a)\ \qquad \mbox{for all }a\in\h,
  \label{coa1} \\[1.4 mm]&& (\cop\oti\id)\circ\cop(\bfe)\cdot\coa =\coa=
  \coa\cdot(\id\oti\cop)\circ\cop(\bfe), \label{coa2} \\[1.4 mm]&&
  \coa\cdot\coa^*=(\cop\oti\id)\circ\cop(\bfe),\qquad
  \coa^*\cdot\coa=(\id\oti\cop)\circ\cop(\bfe).  \label{coa3} \end{eqnarray}

\item The coassociator \coa\ and cocommutator \coc\ satisfy the
compatibility requirements needed to obtain a
braided monoidal rigid $C^*$-category of \h-\rep s. \\
These are the triangle identity
  \be  (\id\oti\cou\oti\id)(\coa) =(\lambda\otim\UN)\cdot\cop(\UN)
  \cdot(\UN\otim\rho^*)\, , \labl{tri}
the square identities
  \be  \apo(\lambda)\cdot\apo(\coa_1)\cdot r\cdot\coa_2\cdot l\cdot
  \apo(\coa_3)\cdot\apo(\rho^*) =\UN= \lambda^*\cdot\coa_1^*\cdot l\cdot
  \apo(\coa_2^*)\cdot r\cdot\coa_3^*\cdot\rho \,, \labl{squ}
the pentagon identity
  \be  (\cop\oti\id\oti\id)(\coa) \cdot (\id\oti\id\oti\cop)(\coa) =
  (\coa\otim\UN)\cdot(\id\oti\cop\oti\id)(\coa) \cdot(\UN\otim\coa)\,,
  \labl{pen}
and the hexagon identities
  \begin{eqnarray} &&\coa_{231}\cdot(\cop\oti\id)(\coc)\cdot\coa_{123}
  =\coc_{13}\cdot\coa_{132}\cdot\coc_{23} \,, \label{hex1} \\[1.5 mm]
  &&\coa_{312}^*\cdot(\id\oti\cop)(\coc)\cdot\coa_{123}^*=\coc_{13}\cdot
  \coa_{213}^*\cdot\coc_{12}\, .  \label{hex2} \end{eqnarray}
Here we used the notation $\coa\equiv\coa_{123}=
\sum_i \coa_{1,i}\otim\coa_{2,i}\otim\coa_{3,i}=
\coa_1\otim\coa_2\otim\coa_3$; similarly, $\coa_{231}:=\coa_2
\otim\coa_3 \otim\coa_1$, $\coc_{13}:=\coc_1\otim\UN\otim\coc_2$, etc.

\item The monodromy matrix $Y$, an element of $M_{\vert\hh\vert}(\H)$
to be defined in \erf{yrs} below, is invertible.
\end{itemize}

The coproduct is not necessarily coassociative and unit preserving,
and hence \rha s are generically not genuine Hopf algebras.
Thus they share a lot of the properties of quasitriangular
quasi-Hopf algebras \cite{drin} and weak quasi-Hopf algebras
\cite{masc,scho,fk}.
The main distinctive features are the ${}^*$-algebra properties and the
further restriction that the monodromy matrix be invertible (the latter is
the only one among the properties that excludes group algebras of finite
non-abelian groups).

The representations of \H\ are ${}^*$-algebra homomorphisms
$D\colon\ \H\to M_n(\complex)\,.$ They are completely reducible. The
defining unitary irreducible representations $D_r,\, r\in\hh$, of \H\ read
  \be  D_r^{ij}(a):=a_r^{ij}\qquad {\rm for}\ \
  a=\sumh p \sumi{i,j}p a_p^{ij}\e pij\in\H \,; \ee
the label corresponding to the co-unit is taken to be 0, i.e. $\cou\equiv D_0$.
Here $a_p^{ij}\in\Co$ are the components of $a\in\h$ in the basis of \h\
that is provided by the matrix units \e rij, $r\in\hh,\, i,j=1,\ldots ,n_r$.
The coproduct \cop\ allows to define the product of representations via
$(D_1\tims D_2)(a):=(D_1\otim D_2)(\cop(a))$.
According to \erf{alr}, products of representations with the \onedim\
\rep\ \cou\ lead to representations that are equivalent to
the original ones. The properties \erf{coc1} -- \erf{coa3} of \coc\ and \coa\
ensure the commutativity and associativity
of the product of representations up to unitary equivalence.
Via the antipode \apo, one defines the contragredient
$\bar D$ of a representation $D$, $\bar D(a):=D^t(\apo(a))$
for all $a\in\H$ (\,$^t$ denotes matrix transposition).
The properties of \apo\ ensure that $\bar D\colon\,\H\to M_n(\complex)$ is a
${}^*$-homomorphism, with $n$ the dimension of $D$, and the elements
$l$ and $r$ serve as natural
$(D\tims\bar D\vert\cou)$ and $(\cou\vert\bar D\times D)$ intertwiners.

The coproduct \cop\ can be viewed as an
embedding of \H\ into $\H\oti\H$. Such embeddings are relevant
only up to inner unitary automorphisms of $\H\oti\H$, because only
unitary equivalence classes of \rep s possess an invariant meaning.
Accordingly, the same vector space \h\ endowed with the same algebra
structure, but with the coproduct
$\cop_U(a):=U\cop(a)U^*$ for some element $U$ of the set
  \be  {\cal U}_2^{}:=\{U\in\H\oti\H \mid UU^*=\UN\otim\UN\} \,, \labl{u2}
is considered to be equivalent as a \rha\ with the original one,
$\H_U\simeq\H$. This definition is meaningful since $\simeq$ is an
equivalence relation, and since by setting
$\rho_U:=\cou(U_1)U_2\rho$, $\lambda_U:=U_1\cou(U_2)\lambda$,
$l_U:=U_1 l\apo(U_2)$, $r_U^{}:=\apo(U_1^*)r U_2^*$,
$\coc_U:=U_{21}\coc U_{12}^*$
(with $U=U_1\otim U_2\equiv \sum_i U_{1,i}\otim U_{2,i}$), as well as
  \be  \coa_U:=U_{12}^{}\,[(\cop\oti\id)(U)]\,\coa\, [(\id\oti\cop)(U^*)]\,
  U_{23}^*, \labl{aU}
the algebra $\H_U\equiv(\H,\cou,\cop_U,\rho_U,\lambda_U,$%
 $\coc_U,\coa_U,\apo,l_U,r_U)$
\,again satisfies the properties of a \rha\ listed above. One can use
this `gauge freedom' to reach a canonical form of \h; in
particular, with a suitable choice of $U$ one can set $\lambda=\UN=\rho$
\,so that one arrives at the properties
  \be  (\cou\oti\id)\circ\cop=\id=(\id\oti\cou)\circ\cop, \ \qquad
  (\id\oti\cou\oti\id)(\coa)=\cop(\UN)  \labl{eid}
of the co-unit and of \coa\ in place of \erf{rar} and \erf{tri}, and
then the relations $(\cou\oti\id\otimes {id})(\coa) =\cop(\UN) =
(\id\oti\id\oti\cou)(\coa)$ and $\cou(\coc_1)\cdot\coc_2=\UN=
\coc_1\cdot\cou(\coc_2)$ are satisfied as well.

Employing another gauge freedom connected to the particular choice of the
contragredient representation within its unitary equivalence
class, $\apo_\U(a):=\U\apo(a)\U^{-1}$, $l_\U:=l\U^{-1}$, $r_\U:=\U r$ for
 $\U\in\h$ with $\U\U^{-1}=\UN$ and $|\U|$ central, one proves that the
antipode and the intertwiners $r$ and $l$ can be chosen as
  \be  \apo(\e pij)=\e{\bar p}ji \labl{u11}
for all $p\in\hh$ and all $i,j=1,\ldots ,n_p$, and
  \be  \chi_p\cdot l\cdot e_p=\apo(r^*) \cdot e_p  \labl{u12}
for all $p\in\hh$, with $\chi_p$ some scalar phase factor. If the sector
labelled by $p$ is non-selfconjugate, then one can in fact fix the
gauge such that $\chi_p=1$, while if the sector is self-conjugate, then
the phase $\chi_p$ takes one of the values $\chi_p=\pm1$, and the value
of this sign is an intrinsic property of the sector \cite {frrs2}. For
$\chi_p=-1$ the self-conjugate sector $p\in\hh$ is called pseudoreal,
while for $\chi_p=1$ it is called real.
 Using the remaining ${\cal U}_2$ gauge freedom, $r$ can be transformed
into a positive invertible central element of $\H$,
  \be  r=\sumh p r_p\,e_p \labl{215}
with $r_p=r_{\bar p}\in\reals_+$ for all $p\in\hh$, while still keeping
the relation \erf{u12}.

To obtain information about the braiding properties of the coproduct,
it is convenient to use amplimorphisms of \H\ instead of
representations, since they can be endowed with an equivalent, but strict
monoidal structure \cite{SV}. Owing to the existence of a left inverse of an
amplimorphism, one can introduce the notion of conditional
expectations, statistics parameters and index.
According to its definition \erf{amp}, an amplimorphism of $\H$ is
a ${}^*$-algebra monomorphism from \h\ to $M_n(\H)$.
The linear space $(\mu\vert\nu)$ of intertwiners between amplimorphisms
$\mu\colon\H\to M_m(\H)$ and $\nu\colon \H\to M_n(\H)$ is
the subspace of all elements $T$ of\, Mat$(m\times n,\H)$ for which
 $\mu(a)T=T\nu(a)$ for all $a\in\H$ and $\mu(\UN)T=T=T\nu(\UN)$.
Amplimorphisms $\nu_1$ and $\nu_2$ are called equivalent, $\nu_1\sim\nu_2$,
if there is a $T\in(\nu_1|\nu_2)$ satisfying
$TT^*=\nu_1(\UN),\;T^*T=\nu_2(\UN).$

One can define subobjects, direct sums and an associative product
$(\mu\times\nu)^{i_1j_1,i_2j_2}(a):=\mu^{i_1i_2}(\nu^{j_1j_2}(a))$
of amplimorphisms. By composing an amplimorphism $\mu\colon H\to M_m(H)$
with the co-unit, it leads to a representation
  $D_\mu:=\cou\circ\mu\colon\ \H\to M_m(\Co),$
  $D^{ij}_\mu(a):=\cou(\mu^{ij}(a))$
for $a\in\H$ and $i,j=1,\ldots,m.$
Conversely, any non-zero representation $D$ of \H\ of dimension $m$
defines a {\em special amplimorphism\/} $\mu_D\colon H\to M_m(H)$ via
  \be  \mu^{}_D:=(\id\oti D)\circ\cop \,. \ee

The braiding of amplimorphisms is described by the statistics operators
\stop. For special amplimorphisms $\mu_1,\, \mu_2$ corresponding to
representations $D_1$ and $D_2$, the statistics operator
$\stopt(\mu_1;\mu_2)$ is an intertwiner between
$(\mu_2\times\mu_1)$ and $(\mu_1\times\mu_2)$, defined as
  \be  \stopt(\mu_1;\mu_2)=[(\id\otim D_2\otim D_1)(\coa) ]\cdot
  [\UN\otim (\tau_{12}\circ( D_1\otim D_2)(\coc))]\cdot
  [(\id\otim D_1\otim D_2)(\coa^*)] \labl{qrq}
($\tau_{12}$ interchanges the tensor product factors of the underlying
\rep\ spaces). The statistics operators are unitary in the sense that
 $\stopt(\mu_1;\mu_2)\cdot\stopt(\mu_1;\mu_2)^*=(\mu_2\times\mu_1)(\bfe)$ and
 $\stopt(\mu_1;\mu_2)^*\cdot\stopt(\mu_1;\mu_2)=(\mu_1\times\mu_2)(\bfe)$,
and they give rise to a \rep\ of coloured braids.

We can restrict our attention to amplimorphisms generated by
representations. Correspondingly,
an amplimorphism $\nu\colon\H\to M_n(\H)$ is called {\em natural\/}
if $\nu\sim\mu_D$, i.e.\ if there exists a representation $D\colon\H\to
M_m(\Co)$ and an equivalence $T\in(\mu_D^{}\vert\nu)$.
Natural amplimorphisms are closed with respect to taking products and
contragredients. For special amplimorphisms with non-zero $D$, there is
a partial isometry $P_\mu \in(\mu_{\bar D}\times\mu_D\vert\id)$ ,
  \be  P_\mu^{ij,\cdot}={1\over \sqrt{{\rm tr}\, D(rr^*)}}\,\coa_1
  \cdot D^{ji}(\coa_3 r^*\apo(\coa_2)),\qquad i,j=1,\ldots,{\rm dim}\,D
  \,, \labl{pij}
which implies a similar partial isometry $P_\nu\in(\bar\nu\times\nu\vert\id)$
for natural amplimorphisms $\nu\sim\mu_D$.
As a consequence of the so-called initial conditions \cite V, the
statistics operators for a natural amplimorphism $\nu$ and the identity read
  \be  \stopt(\id;\nu)=\nu(\UN)=\stopt(\nu;\id) \,. \labl{nu0}
A {\em standard left inverse\/}
$\Phi_\nu\colon\, M_n(\H)\to \H$ of a natural amplimorphism
$\nu\colon\, \H\to M_n(\H)$ (with $\nu\sim\mu_D$ for a non-zero
$D$) is then  defined as
  \be  \Phi_\nu(A):=P^*_\nu\cdot\bar \nu(A)\cdot P_\nu \qquad
  \mbox{for all }\ A\in M_n(\H) \,. \labl{pnp}
$\Phi_\nu$ is a unit preserving positive linear map satisfying
$\Phi_\nu(\nu(a)\cdot B\cdot\nu(c))=a\cdot\Phi_\nu(B)\cdot c$
for all $a,c\in\H$ and all $B\in M_n(\H)$.

The {\em statistics parameter matrix\/} $\Lambda_\nu\in M_n(\H)$ and the
{\em statistics parameter\/} $\lambda_\nu\in\H$ of a natural
amplimorphism $\nu\colon\, \H\to M_n(\H)$ are defined as
  \be  \Lambda_\nu:=\Phi_\nu(\stop_\nu),\qquad\
  \lambda_\nu:=\Phi_\nu(\Lambda_\nu) \,, \labl{ll}
where $\stop_\nu\equiv\stop(\nu;\nu)$.
The statistics parameter $\lambda_\nu$ is an element of the
center of \H\ and depends only on the equivalence class of $\nu$. For an
irreducible amplimorphism $\nu_r,\, r\in\hh$, $\Lambda_{\nu_r}$ takes the form
  \be  \Lambda_{\nu_r}^{}={\omega_r\over d_r}\cdot\nu_r(\UN) \,,  \labl W
so that
  \be  \lambda_r={\omega_r\over d_r}\cdot \UN \,; \labl{wde}
the pure phase $\omega_r$ is called the {\em statistics phase},
and the positive real number $d_r$ the {\em statistical dimension\/}
of the irreducible representation $D_r$.

 Finally, the monodromy matrix $Y\in M_{\vert\hh\vert}(\H)$ is defined by
  \be  Y_{rs}:=d_rd_s\cdot\Phi_r\Phi_s(\stop(\nu_r;\nu_s)\cdot
  \stop(\nu_s;\nu_r)), \qquad r,s\in\hh \,.  \labl{yrs}
One can show that $Y_{rs}=y_{rs}\cdot\UN$ with $y_{rs}\in\Co$, and if
$Y$ is invertible, as is required for \rha s, then similarly to
\cite{rehr4} one proves that
  \be  V(\MS)_{rs}:={1\over \vert\sigmA\vert}\cdot y_{rs},\qquad
  V(\MT)_{rs}:=\left({\sigmA\over\vert\sigmA\vert}\right)^{1/3}\cdot
  \delta_{rs}\,\omega_r, \, \label{mod} \ee
with
  \be  \sigmA := \sumh r d_r^2\,\omega_r^{-1} \,, \labl{sig}
provides a unitary representation $V$ of the modular group $\Gamma$.
The statistics parameters and the monodromy matrix are
independent of the gauge choices described above, while the statistics
operators are invariant up to unitary equivalence.
The invertibility of the monodromy matrix requires that
$\sigmA$ as defined in \erf{sig} satisfies
  \be  |\sigmA|^2=\sigmASQ  \,, \labl s
where
  \be  \sigmASQ\,:=\sumh r d_r^2   \labl{s'}
(in analogy with the theory of finite groups, we may call $\sigmASQ$ the
{\em order\/} of the \rha\ \h).
If this equality holds, then the number
  \be  c=\frac{4\ii}\pi \cdot\log\frac\sigmA{|\sigmA|}\,\in[0,8)  \labl c
plays the role of the `central charge' $c$ of \h, which should be
equal (mod 8) to the Virasoro central charge of any
\cft\ model that possesses \h\ as its \qs.
If \erf s is not fulfilled so that the monodromy matrix is degenerate,
then the algebra \h\ is said to be degenerate, too.
(Dropping the requirement of invertibility of the monodromy
matrix, one can study degenerate \RHA s; they can correspond to field
theories with finite number of superselection sectors among which not
only the vacuum sector is degenerate \cite{rehr4}.)

\sect{$F$ and $R$ matrices and polynomial equations}

\subsection{\CGC s}

The coproduct can be described conveniently in terms of \CG type
coefficients $\cgc pqrijk\alpha$, where $p,q,r\in\hh$,
$\alpha\in\{1,2,...\,,\npqr\}$, and $i\in\{1,2,...\,,n_p\}$,
etc. To this end, first note that a simple matrix \alg\ $M_n$ has a
single \irrep\ acting on the $n$-dimensional Hilbert space $V_n\cong
\complex^n$. Therefore one can identify $M_n$ with $V_n\oti V_n^*$, in
such a way that the matrix units $e^{i,j}$ correspond to $v^i\otim v^{*j}$
with $\{v^i\}$ an orthonormal basis of $V_n$ and $\{v^{*i}\}$ the dual
basis of $V_n^*$. The coproduct then corresponds to a map
  \be  v_p^i\mapsto\sumn\alpha pqr \sumi jq \sumi kr \cgc qrpjki\alpha
  v_q^j\otim v_r^k \,, \qquad  v_p^{*i}\mapsto\sumn\alpha pqr \sumi jq
  \sumi kr \cgcs qrpjki\alpha v_q^{*j}\otim v_r^{*k} \,. \ee
Thus on matrix units, the coproduct \cop\ acts as
  \be  \cop(\etij{}) = \sumh{r,s} \sumi{k,l}r \sumi{k',l'}s
  \sumn\alpha rst \cgc rstk{k'}i\alpha \cgcs rstl{l'}j\alpha
  \,\e{rs}{kk'}{ll'} \,. \labl{cop}
Here and below we use the notation
  \be  \e {p_1p_2...p_m} {i_1i_2...i_m} {j_1j_2...j_m}
  \equiv \epij1\otim \epij2\otim\ldots\otim \epij m \ee
for the matrix units of $\h^{\otimes m}$.  The fact that \cop\ is an
algebra homomorphism implies the orthogonality property
  \be  \sumi ir \sumi js \cgcs rsuijk\alpha \cgc rsvijl\beta
  = \delta_{uv}\delta_{kl}\delta_{\alpha\beta}  \labl{ort}
of \CGC s.

The opposite coproduct \copp\ can be decomposed analogously; it is then
just given by \erf{cop} with $\e{rs}{kk'}{ll'}$ replaced by
$\e{sr}{k'k}{l'l}$. The cocommutator \coc\ intertwines \cop\ and
\copp; thus it can be written as
  \be  \coc = \sumh{r,s,t} \sumn{\alpha,\beta}rst
  \R rst\alpha\beta \sumi{i,j}r \sumi{i',j'}s \sumi kt
  \cgc srt{i'}ik\alpha \cgcs rstj{j'}k\beta \, \e{rs}{ii'}{jj'} \, \labl{coc}
with $\R pqt\alpha\beta\in\complex$.
Because of the relation \erf{coc3}, for fixed $p$, $q$ and $t$,
\R pqt\alpha\beta\ is a unitary matrix in the indices $\alpha$ and $\beta$.
Similarly, one may insert the decomposition \erf{cop}
in the combinations $(\id\oti\cop)\circ\cop$ and $(\cop\oti\id)\circ\cop$.
The coassociator intertwines between these, and therefore
its most general form reads
  \be  \begin{array}{l} \displaystyle{ \coa = \sumh{p,q,r} \sumh{t,u,v}
  \sumn\alpha pqu \sumn\beta urt \sumn\gamma qrv \sumn\delta vpt
  \F{pqr}\alpha u\beta \gamma v\delta t }\\ \mbox{}\\[-2 mm]
  \hsp{2} \displaystyle{
  \cdot \sumi{i,j}p \sumi{i',j'}q \sumi{i'',j''}r \sumi kt \sumi lu \sumi mv
  \cgc pqu i{i'}l\alpha \cgc urt l{i''}k\beta
  \cgcs qrv {j'}{j''}m\gamma \cgcs vptjmk\delta    \,
  \e{pqr}{ii'i''}{jj'j''} \, }  \end{array} \labl{coa}
with $\Fpqr\in\complex$.
{}From the requirements \erf{coa3} it follows that for fixed $p$, $q$,
$r$ and $t$ for which $\FF pqr\cdot\cdot t$ are non-vanishing, \Fpqr\
is a unitary matrix in the (multi-)indices $(\alpha,u,\beta)$ and
$(\gamma,v,\delta)$, i.e.
  \be  \sumh w \sumn \mu qrw \sumn \nu pwt \F{pqr}\alpha u\beta\mu w\nu t
  \Fb{pqr}\gamma v\delta\mu w\nu t=\delta_{\alpha\gamma}\delta_{\beta\delta}
  \delta_{uv}  \labl u
for $N_{pq}^{\ \,u}N_{ur}^{\ \,t}>0$ and $\alpha\in\{1,2,...\,,N_{pq}
^{\ \,u}\}$, $\beta\in\{1,2,...\,,N_{ur}^{\ \,t}\}$.

We also note that the property \erf{eid} of the co-unit implies that
for nonzero $F$-coefficients we have
  \be  1=\F{pqr}{}\cdot{}{}\cdot{}t\in M_1, \ee
whenever $p,\,q$ or $r$ equals $0\in\hh$.

\subsection{A natural gauge choice}

The explicit form of the \CGC s depends on the gauge choice mentioned
in the previous section, and therefore contains a lot of redundant
information. As we will now show, there exists a natural gauge choice
in which the \CGC s take an extremely simple form.
We start by observing that the maximal abelian subalgebras ($A$) and
the centers ($Z$) of $H,\, H\oti H$ and $H\oti H\oti H$ are given by
  \be  \{ \e rii\},\quad\{ \e rii\otim \e sjj \},\quad
  \{ \e rii\otim \e sjj\otim \e tkk \} \ee
and by
  \be  \{ e_r\},\quad\{ e_r\otimes e_s\},\quad\{ e_r\otimes e_s\otimes e_t\}
  \,, \ee
respectively, where $r,s,t\in\hh$, $i=1,2,\dots,n_r$ etc., and
$e_r\equiv\sumi ir \e rii$. In other words, they are spanned by the
tensor products of minimal and central projections, respectively.

There is a \,${\cal U}_2$ gauge of the type \erf{u2} for which
the coproduct maps $A(H)$ into $A(H)\otimes A(H)$.
(For the centers, the analogous statement does not hold in general.)
This motivates to introduce the notations
  \be \begin{array}{l} a(rs;1,2):=(e_r\otimes e_s)(a^{(1)}\otimes a^{(2)})
   \,,\\[2.2 mm] a(rst;11,12,2):=(e_r\otimes e_s\otimes e_t)
  (a^{(11)}\otimes a^{(12)}\otimes a^{(2)}) \,, \end{array} \ee
etc., for all $a\in H$. The intertwiner properties of $R$ and $\varphi$
imply that
  \be  \e pii(rs;1,2) \sim \e pii(rs;2,1) \ee
and
  \be  \e pii(rst;11,12,2) \sim \e pii(rst;1,21,22) \ee
are equivalent projections in $H\otimes H$ and $H\otimes H\otimes
H$, respectively. We will see that these equivalences
imply the symmetry and associativity of the fusion coefficients
\npqr; in the present context, these are defined by the product
decomposition
  \be  D_p\times D_q =\bigoplus_{r\in\hh}\npqr D_r  \ee
of irreducible \h-\rep s. In other words, the set $\hh$ can be
identified with the \rep\ ring of the algebra \h, and the fusion
coefficients are the structure constants of this ring.

Moreover, the gauge can be chosen such that
  \be  \e pii(rs;1,2)=\e pii(rs;2,1) \quad {\rm for}\quad r\not=s
  \, .\labl{3.1}
Since by the coproduct projections are mapped to projections,
the gauge choice implies that the images of the minimal
projections $\e tii\in A(H)$ are linear combinations of
minimal projections $\e rjj\otim\e skk\in A(H)\otim
A(H)$ with coefficients 0 or 1:
  \be (e_r\otimes e_s)\cdot\cop(\e tii)\equiv\e tii(rs;1,2)=
  \sumn \alpha rst \e tii(rs,\alpha;1,2)=
  \sumn \alpha rst \e r{j^\alpha_{ti}} {j^\alpha_{ti}}\otim
  \e s{k^\alpha_{ti}} {k^\alpha_{ti}} \,.\labl{3.2}
Here the notations
  \be  j^\alpha_{ti} \equiv j(r,s;\alpha,t,i)\in\{1,\ldots ,n_r\}\,, \quad
  k^\alpha_{ti} \equiv k(r,s;\alpha,t,i)\in\{1,\ldots ,n_s\} \labl{jk}
were used for short.  In terms of \CGC s, \erf{3.2} corresponds to the choice
  \be \cgc rstjki\alpha =
  \delta^{}_{j,j_{ti}^\alpha}\delta^{}_{k,k_{ti}^\alpha} \,. \labl{c1}
Owing to the orthogonality of the minimal projections \e tii,
the index pairs $(j^\alpha_{ti},k^\alpha_{ti})$ are
different for different combinations of $t,i,\alpha$ values. The
fusion coefficients \nrst\ coincide with the number of
one-dimensional projections in $\e tii(rs;1,2)$; thus
  \be  \nrst={\rm tr}\, \e tii(rs;1,2)={\rm tr}\, \e tii(rs;2,1)=
  {\rm tr}\, \e tii(sr;1,2)=N_{sr}^{\ \,t} \,, \labl{3.3}
which proves symmetry of the fusion coefficients.

Due to the properties of the coassociator $\varphi$, the equivalence
  \be  \e tii(prs;11,12,2) \sim  \e tii(prs;1,21,22)  \labl A
of projections holds as well. Using the short hand notation
  \be  r_i\equiv \e rii  \ee
for $r\in\hh$ and $i=1,2,...\,,n_r$, one has
  \begin{eqnarray} \e tii(prs;11,12,2)
  &=&(e_p\otiM e_r\otiM e_s)\cdot(\cop\oti\id)\Delta(t_i)
  \nonumber\\
  &=&\sum_q\sumn\alpha qst (e_p\otimes e_r)\cdot
  \cop(q_{j^\alpha_{ti}})\otimes s_{k^\alpha_{ti}}
  =\sum_q\sumn\alpha qst \sumn\beta prq p^{}_{j^\beta_{qj^\alpha_{ti}}}\otiM
  r^{}_{k^\beta_{qj^\alpha_{ti}}}\otiM s^{}_{k^\alpha_{ti}}, \qquad{}
  \label{3.4a}\\[1.1 mm]
  \e tii(prs;1,21,22) &=&(e_p\otiM e_r\otiM e_s)\cdot(\id\oti\cop)\cop(t_i)
  \nonumber\\
  &=&\sum_q\sumn\alpha pqt p^{}_{j^\alpha_{ti}}\otiM
  (e_r\otiM e_s)\cdot\cop(q^{}_{k^\alpha_{ti}})
  =\sum_q\sumn\alpha pqt \sumn\beta rsq
  p^{}_{j^\alpha_{ti}}\otiM r^{}_{j^\beta_{qk^\alpha_{ti}}}\otiM
  s^{}_{k^\beta_{qk^\alpha_{ti}}}. \qquad{}
  \label{3.4b} \end{eqnarray}
Due to our specific gauge choice, the right hand sides of the relations
\erf{3.4a} and \erf{3.4b} are sums
of commuting minimal projections in $A(H)\otimes A(H)\otimes A(H)$.
As a consequence, they must be all
different in order for the sum to be a projection. The equivalence \erf A
of these projections then implies that they have the same rank:
  \be  \sumh q N_{qs}^{\ \,t}N_{pr}^{\ \,q}=\sumh q N_{pq}^{\ \,t}
  N_{rs}^{\ \,q} \,. \labl{3.5}
This proves associativity of the fusion coefficients.
Similarly, by using intertwiners connected to the antipode, one can
see that $\npqr=N_{p\bar r}^{\ \,\bar q}=N_{\bar p\bar q}^{\ \,\bar r}$,
and from the properties of the left inverses it follows that
$d_{\bar p}=d_p$ and $d_p d_q=\sumh r\npqr\,d_r$.
Note that the properties of the fusion coefficients just derived are
gauge independent, since the traces are invariant under ${\cal U}_2$
transformations of the type \erf{u2}.

Once the images $\e tii(rs;1,2)$ are fixed, there is a canonical
choice for the images of the non-diagonal matrix units $\e ti{i'}$, namely
  \be \e ti{i'}(rs;1,2)= \sumn\alpha rst e_r^{j^\alpha_{ti},j^\alpha_{ti'}}
  \otimes e_s^{k^\alpha_{ti},k^\alpha_{ti'}}. \ee
Therefore all information about the coproduct can be coded in the image
of $A(H)$. To visualize the structure of the coproduct one can introduce
the following $N\times N$ matrix $M_\Delta$, where $N=\sum_{r\in\hh} n_r$: the
matrix elements are minimal projections $t_i\equiv\e tii\in A(H)$,
and the columns and rows of $M_\Delta$ are indexed by minimal
projections in $A(H)$ as well; thus the element $t_i$
sitting in the $r_j$th row and $s_k$th column
indicates that $\cop(t_i)$ contains the minimal projection
$r_j\otimes s_k\in A(H)\otimes A(H)$. For instance, in the examples of
the \LY and the Ising fusion rules (compare \erf{ly} and \erf{is} below),
the matrix $M_\Delta$ looks like
  \be  \copmatabb {1_1\\0\\1_1} {1_2\\ \\1_2} \quad \raisebox{-1.7em}{\rm and}
  \quad\ \copmatabcc{1\\0\\2_1\\2_2} {2_1\\2_1\\0\\{~} } {2_2\\2_2\\ \\1}
  \raisebox{-1.7em}, \ee
respectively. The presence of empty entries in these matrices expresses
the fact that the corresponding coproducts are not unit preserving.
Transposition of $M_\Delta$ amounts to considering the opposite coproduct. Thus
if $M_\Delta$ is symmetric, as in the Ising case, then the coproduct is
cocommutative. In the \LY case $M_\Delta$ is not symmetric, and hence
the coproduct is not cocommutative.

\subsection{The pentagon and hexagon equations}

Let us now rewrite the pentagon identity \erf{pen} in terms of the
$F$-matrices that were defined in subsection 3.1. After
inserting the expansion \erf{coa} into \erf{pen}
and then using the orthogonality relation \erf{ort}
(three times on the left hand side and four times on the right hand side),
the pentagon identity reads as
  \be  \sumn\sigma uvt \Fa{pqv}\sigma u\alpha \mu y\beta  t
                 \cdot \Fa{urs}\nu x\delta \sigma v\gamma t
  = \sumh w \sumn\kappa wsy \sumn\lambda pwx \sumn\eta qrw
                       \Fa{pqr}\delta u\alpha \lambda w\eta x
                 \cdot \Fa{pws}\nu x\lambda \mu y\kappa t
                 \cdot \Fa{qrs}\kappa w \eta \beta v\gamma y \,. \labl{penf}
Similarly, inserting the expansions \erf{coc} and \erf{coa} into the
hexagon identities \erf{hex1}, \erf{hex2}, one obtains
  \be \begin{array}{l}
  {\dstyle \sumh u \sumn{\delta,\lambda}urt \sumn\kappa pqu }\,
  \F{rpq}\beta w\gamma\kappa u\delta t \cdot \R urt\delta\lambda \cdot
  \F{pqr}\kappa u\lambda\mu v\nu t= {\dstyle\sumn\alpha prw\sumn\delta qrv }\,
  \R prw\beta\alpha \cdot \F{prq}\alpha w\gamma\delta v\nu t \cdot
  \R qrv\delta\mu \,,
  \\{} \\[-2.1 mm]
  {\dstyle \sumh u \sumn{\delta,\lambda}urt \sumn\kappa pqu }\,
  \Fb{rpq}\beta w\gamma\kappa u\delta t \cdot \R rut\lambda\delta \cdot
  \Fb{pqr}\kappa u\lambda\mu v\nu t= {\dstyle\sumn\alpha prw\sumn\delta qrv}\,
  \R rpw\alpha\beta \cdot \Fb{prq}\alpha w\gamma\delta v\nu t \cdot
  \R rqv\mu\delta \,.
  \end{array} \labl{hexf}

\subsection{The statistics operators and statistics parameters}

Inserting the expansions \erf{coc} and \erf{coa} into the
formula \erf{qrq} for the statistics operators, one obtains
  \be  \begin{array}{l}  \stopt_{qr}^{} = {\dstyle\sumh{p,s,t,u,v}
  \sum_{\alpha,\beta,\gamma,\delta,\atop \lambda,\mu,\nu}}
  \F{prq}\alpha u\beta\gamma s\delta t
  \Fb{pqr}\lambda v\mu\nu s\delta t \R qrs\gamma\nu \\[2.2 mm]
  \hsp{3} \cdot \dstyle{\sumi{i,i'}p \sumi{j,j'}q \sumi{k,k'}r \sumi lt
  \sumi mu \sumi nv}
  \cgc pruikm\alpha \cgc uqtm{j'}l\beta \cgcs pqv{i'}jn\lambda
  \cgcs vrtn{k'}l\mu \, [\e pi{i'}]_{}^{kj',jk'} \,. \end{array}\labl{ep}
Here and below, the \CGC s are to be interpreted according to \erf{c1}.

Also, with the gauge choice \erf{c1}, the intertwiners $r$ and $l$ of \h\
that appear in \erf{alr} are of the form
  \be  r=\sumh p r_p^{}\, \e p11, \qquad l=\sumh p l_p\, \e p11 \labl{rl}
with $r_p\,,l_p\in\complex$, and without loss of generality one can choose
$r_p\in\reals$.
 \futnote{Note that in the natural gauge $r$ and $l$ are generically not
central elements, unlike in the gauge referred to in \erf{215}}.
 The square identities \erf{squ} then reduce to
  \be  r_p l_p\, \F{\bp p\bp}{}0{}{}0{}\bp =1
  = r_p l_p\, \Fb{p\bp p}{}0{}{}0{}p   \labl{sqi}
for all $p\in\hh$. Using this result, it is straightforward to see that
the formula \erf{pnp} for the left inverses leads to
 \futnote{For better readability, multiplicity indices $\alpha,\,\beta
\ldots\,$ that can take only a single value are not written explicitly.}
  \be  \Phi_p([\e qkl]_{}^{m,n})= \sumh s \sum_{\alpha,\beta,\gamma}
  \Fb{s\bp p}\alpha q\beta{}0{}s \F{s\bp p}\alpha q\gamma{}0{}s
  \, \sumi{i,j}s \cgcs qpskmi\beta \cgc qpslnj\gamma \e sij  \,.
  \labl{ph}
That this is indeed a left inverse to $\mu_p$ is easily verified
with the help of the unitarity \erf u of the $F$ matrices.

Combining \erf{ep} and \erf{ph}, one obtains the statistics
parameter matrices in the form
  \be  \begin{array}{l}  \Lambda_p={\dstyle\sumh{q,r,s,t}
  \sum_{\alpha,\beta,\gamma,\delta,\atop \kappa,\lambda,\mu,\nu}}
  \R ppr\mu\nu \Fb{s\bp p}\gamma q\kappa{}0{}s
  \F{s\bp p}\gamma q\lambda{}0{}s
  \Fb{qpp}\lambda s\beta\nu r\delta t \F{qpp}\kappa s\alpha\mu r\delta t
  \\[-1.5 mm] \hsp{12} \cdot {\dstyle \sumi{i,j}s \sumi{k,l}p \sumi mt}
  \cgc sptikm\alpha \cgcs sptjlm\beta [\e sij]_{}^{k,l} \,.
  \end{array} \labl{L0}
Applying to this result the hexagon identity, then the pentagon identity,
and then once more the hexagon identity, it can be rewritten as
  \be  \Lambda_p= {\dstyle\sumh{s,t} \sum_{\alpha,\beta,\gamma}}
  \RRb \bp p0 \F{t\bp p}\gamma s\alpha{}0{}t \Fb{sp\bp}\beta t\gamma {}0{}s
  \; {\dstyle\sumi{i,j}s \sumi{k,l}p \sumi mt}
  \cgc sptikm\alpha \cgcs sptjlm\beta [\e sij]_{}^{k,l} \,.
  \labl L
This expression can be analyzed further by employing the identities \cite{SV}
  \be  \begin{array}{l} \apo(\coc_1^*)\cdot r\cdot\coc_2^*=\zz\cdot l^* \,,
  \\[1.8 mm]  \cop(\UN)=\coa_1^*\coa_1^{(1)} \otim \coa_3^{}r^*\apo(\coa_2^{})
  \apo(\coa_3^*)l^*\coa_2^*\coa_1^{(2)} \,, \end{array} \ee
where $\zz=\sum_p\zz_p e_p$ is an invertible central element of \h.
In the natural gauge, these identities read
  \be  \RRb\bp p0 = \zz_p\bar l_p/r_p \ee
and
  \be  \begin{array}{l}
  {\dstyle \sumh{p,s,t}\sumn\alpha spt \sumi{i,j}s \sumi{k,l}p \sumi mt}
  \cgc sptikm\alpha \cgcs sptjlm\alpha \e{sp}{ik}{jl} \\[1 mm]
  \hsp{3.2} = {\dstyle \sumh{p,s,t} \sumn{\alpha,\beta}spt\sumn\gamma t\bp s}
  \, \Fb{sp\bp}\beta t\gamma{}0{}s \F{t\bp p}\gamma s\alpha{}0{}t \,
  \bar r_p \bar l_p \, {\dstyle \sumi{i,j}s \sumi{k,l}p \sumi mt}
  \cgc sptikm\alpha \cgcs sptjlm\beta \e{sp}{ik}{jl} \,. \end{array}\ee
Combining these results with \erf L, we obtain
  \be  \bigoplus_{p\in\hh}\mu_p^{}(\UN) \equiv
  (\id\oti\bigoplus_{p\in\hh}D_p)(\cop(\UN)) =
  \bigoplus_{p\in\hh}|r_p|^2 \zz_p^{-1} \Lambda_p \,. \labl;
This shows, in particular, that $\Lambda_q$ is a
scalar multiple of $\mu_q(\UN)$, in accordance with \erf W.

By application of the left inverse $\Phi_q$ to \erf; one then learns by
comparison with \erf{ll} that the statistics parameter $\lambda_q$ must be
a scalar multiple of \UN, in agreement with the general result \erf{wde}.
On the other hand, acting by $\Phi_p$ directly on $\Lambda_p$
as given in \erf L, one finds that the statistics parameters satisfy
  \be  \lambda_p=\sumh{s,t} \sum_{\alpha,\beta,\gamma,\delta}
  \RRb \bp p0 \F{t\bp p}\gamma s\alpha{}0{}t \Fb{t\bp p}\delta s\alpha {}0{}t
  \F{t\bp p}\delta s\beta {}0{}t \Fb{sp\bp}\beta t\gamma {}0{}s
  \, e_t  \,. \labl l
In this formula, it is not manifest that $\lambda_p$ obeys \erf{wde}.
Conversely, by implementing \erf{wde} we conclude that the prefactor of $e_t$
in \erf l is in fact independent of the label $t$. Thus for any choice
of $t$ this prefactor equals its value at $t=0$, so that \erf l reduces to
  \be  \lambda_p=\UN\cdot \RRb \bp p0 \Fb{\bp p\bp}{}0{}{}0{}\bp
  =\UN\cdot \RRb \bp p0  /\bar r_p \bar l_p \,. \labl-

\subsection{The monodromy matrix}

Inserting the previous results into the definition \erf{yrs} of the
monodromy matrix, analogous computations as before lead to
  \be  Y_{pq}=d_p d_q \!\!{\dstyle \sumh{r,s,t,u} \sum_{\alpha,\beta,
  \gamma,\delta,\kappa, \atop \mu,\nu,\rho,\eta,\zeta}} \!\!
  \F{rqp}\alpha u\beta\gamma s\delta t \Fb{rqp}\kappa u\rho\eta s\delta t
  \Fb{u\bq q}\zeta r\alpha{}0{}u \F{u\bq q}\zeta r\kappa{}0{}u
  \Fb{t\bp p}\mu u\beta{}0{}t \F{t\bp p}\mu u\rho{}0{}t
  \R pqs\gamma\nu \R qps\nu\eta \,e_t \,.  \labl Y
Again, this must in fact be proportional to \UN, i.e.\ the prefactor of $e_t$
must be independent of $t$ and hence equal to its value for $t=0$. Thus
we can simplify \erf Y to
  \be  Y_{pq}=\UN\cdot d_p d_q {\dstyle \sumh r \sum_{\alpha,\beta,
  \gamma,\delta,\mu,\nu}} \!
  \F{rqp}\alpha\bp{}\beta{\bar r}{}0 \Fb{rqp}\gamma\bp{}\delta{\bar r}{}0
  \Fb{\bp\bq q}\mu r\alpha{}0{}\bp \F{\bp\bq q}\mu r\gamma{}0{}\bp
  \R pq{\bar r}\beta\nu \R qp{\bar r}\nu\delta \,.  \labl z
This implies in particular that
  \be  Y_{p0}=Y_{0p}=d_p\,\UN. \ee

\sect{Paths and tetrahedra}

\subsection{Path Hilbert spaces and intertwiners}

Denote by $\HS=\bigoplus_{r\in\hh} \HS_r,\, \vert\HS_r\vert=n_r$, the \rep\
space of the universal representation $D=\bigoplus_{r\in\hh} D_r$ of $H$.
Then we can introduce length $l$ path Hilbert spaces
  \be \begin{array}{cl} \HS(1)\equiv \HS, &l=1,\\[2.1 mm]
           \HS(1,2),\ \HS(2,1), &l=2,\\[2.1 mm]
  \HS(11,12,2),\ \HS(1,21,22),\ \HS(12,11,2), \ldots , &l=3, \end{array} \ee
etc., as subspaces of the $l$-fold tensor product
$\HS\otimes \HS\otimes \ldots\otimes \HS$ corresponding to the
range of the projections $\UN,\,\cop(\UN),\,\copp(\UN),\,
(\cop\oti\id)\circ\cop(\UN),$ etc. Obviously, path
Hilbert spaces with equal length $l$ have the same dimension, e.g.
  \be  \begin{array}{l} \vert \HS(1)\vert =
  {\displaystyle \sum_{p\in\hat H}} n_p \,,\qquad
  \vert \HS(1,2)\vert=\vert \HS(2,1)\vert =
  {\displaystyle \sum_{p,r,s\in\hat H}} n_pN_{rs}^{\ \,p} \,, \\{~}\\[-2.8 mm]
  \vert \HS(11,12,2)\vert=\vert \HS(1,21,22)\vert=\vert \HS(12,11,2)\vert=
  \dots = {\displaystyle \sum_{p,r,s,q,t\in\hat H}} n_pN_{rs}^{\ \,p}
  N_{qt}^{\ \,r} \,,\end{array} \ee
and so on. Now the elements $\PA$ of a path Hilbert space are defined as
unit vectors in the range of one-dimensional subprojections occuring
in the corresponding images
  \be  \e tii(1),\ \e tii(1,2),\ \e tii(2,1),\ \e tii(11,12,2),\
  \e tii(1,21,22),\ \ldots, \qquad t\in\hat H,\ i=1,\ldots, n_t \,, \ee
of minimal projections $\e tii\in A(H)$.
Hence the paths form an orthonormal basis in the corresponding
path Hilbert space, and a path $\PA\in\HS(\cdo)$ is completely prescribed
if we characterize the rank one subprojections of
$\e tii(\cdo),\, t\in\hat H,\, i=1,\ldots, n_t$. For example,
  \be  \begin{array}{ll} \e tii(11,12,2)&=
  {\displaystyle \bigoplus_{q,p,r,s\in\hat H}\,
  \bigoplus_{\alpha=1}^{N_{qs}^{\ t}}\bigoplus_{\beta=1}^{N_{pr}^{\ q}}}\
  p_{j^\beta_{q\,j^\alpha_{ti}}}\otimes
  r_{k^\beta_{q\,j^\alpha_{ti}}}\otimes s_{k^\alpha_{ti}}\\[1.7 mm] &=
  {\displaystyle \bigoplus_{q,p,r,s\in\hat H}\,
  \bigoplus_{\alpha=1}^{N_{qs}^{\ t}}\bigoplus_{\beta=1}^{N_{pr}^{\ q}}}\
  \vert\PL(ti;\alpha_{qs}^t;\beta_{pr}^q)\rangle\;
  \langle\PR(ti;\alpha_{qs}^t;\beta_{pr}^q)\vert\, .\end{array} \ee
We refer to the pair $\sigma(\PA)=(t,i)$ as the initial point or
{\em source\/} of a path $\PA=\PA(ti;\alpha_{qs}^t;\ldots)$.
The other $l-1$ arguments $\,\alpha_{qs}^t, \dots\;$ of a length $l$ path
are called {\em vertices}. The final point or {\em range\/}
of a path $\PA$ is an $l$-tuple $(r_1,i_1;\ldots;r_l,i_l)$ of pairs,
where $r_k\in\hat H$ and $i_k=1,\ldots, n_{r_k}$ for $ k=1,\ldots,l$.
The labels $r_1,\ldots,r_l$ are those
labels of the vertices of $\PA$ that do not occur twice irrespective
of the label corresponding to the initial point (except, of course,
for length 1 paths, where the range coincides with the source).
 \futnote{The order of these labels is obtained if we arrange
the vertices in a down and left or right going path in a
pyramid determined by the top (being the first vertex that has a common label
with the source) and by writing subsequently the next
vertex in $\PA$ down left or down right depending on their common index.
Continuing any label which does not occur twice in the vertices to the
bottom of the pyramid then it is the bottom row that gives the
ordering of the indices $r_k$.}
 The indices $i$ and $i_1,\ldots,i_l$ of the source and the
range of a path will be called {\em flags}. The range flags do not
appear explicitly in the description ot a path $\PA$, since because
of \erf{c1} they are uniquely determined by the source
and the vertices of $\PA$.

To make contact to the formulation in terms of the matrices $R$
and $F$ introduced in \erf{coc} and \erf{coa} above, we observe that
the cocommutator \coc\ and the coassociator \coa\ are
isometries between certain path Hilbert spaces of the same length
 \futnot{They are not unitaries since they map different Hilbert
spaces into each other, and they are not just partial isometries because
their initial and final projections are the whole path Hilbert spaces.}
 and examine the matrix elements of these isometries in the path basis. The
smallest path Hilbert spaces where they are defined are
  $ R\colon\ \HS(1,2)\to\HS(2,1)$ and $
  \varphi\colon\ \HS(1,21,22)\to\HS(11,12,2)$, respectively.
The images of matrix units in the path basis can be written as (say)
  \be  \e tij(11,12,2)=\sum_{p,q,r,s\in\hat H}\,
  \sumn\alpha qst \sumn\beta prq\
  \vert\PL(ti;\alpha_{qs}^t;\beta_{pr}^q)\rangle\;
  \langle\PR(tj;\alpha_{qs}^t;\beta_{pr}^q)\vert \,,  \ee
that is, only the source is different (if $i\not= j$) in every term in the
sum. Therefore the intertwiner property
  $\e tij(11,12,2)\cdot\varphi=\varphi\cdot \e tij(1,21,22)$
  (for all $t\in\hat H$ and all $i,j=1,\ldots, n_t$)
of $\varphi$ and the orthogonality of the minimal projections of $H$ imply the
vanishing $\langle\PL(ti;\alpha_{qs}^t;\beta_{pr}^q)\vert\,\coa\,\vert
  \PR(t'i'; \tilde\alpha_{p'q'}^{t'};$ $\tilde\beta_{r's'}^{q'})\rangle=0$
\,of the matrix elements between $\PL\in\HS(11,12,2)$ and $\PR\in\HS(1,21,22)$,
unless $t=t',i=i',p=p',r=r',s=s'$. This implies that \coa, and analogously
\coc, is a direct sum of isometries,
  \be  \varphi= \bigoplus_{p,q,r\in\hat H} \;\Llb\, \bigoplus_{s\in\hat H}
  \,\bigoplus_{i=1}^{n_s}\, \varphi(si;pqr) \Lrb \,,\qquad
  R= \bigoplus_{p,q\in\hat H}\,\Llb\, \bigoplus_{s\in\hat H}\,
  \bigoplus_{i=1}^{n_s}\, R(si;pq) \Lrb \,, \labl{3.6a}
and we can write
  \begin{eqnarray} \varphi(si,pqr)&=& \!\!\!\sum_{\alpha,\beta,\gamma,\delta}
  \!\vert \PL(si; \beta_{ur}^s, \alpha_{pq}^u)\rangle\,
  \F {pqr}\alpha u\beta \gamma v\delta s \,
  \langle \PR(si; \delta_{pv}^{s}, \gamma_{qr}^v) \vert \,, \label{33}
  \\[1 mm] R(si,pq)&=& \sum_{\alpha,\beta}\, \vert
  \PL(si;\stackrel\circ\alpha\!{}_{pq}^{s}) \rangle\, \R pqs\alpha\beta \,
  \langle \PR(si;\beta_{pq}^s)\vert \,, \end{eqnarray}
where the circle above the vertex of the dual path in $R$ indicates
that this vertex corresponds to the opposite coproduct.
The matrices \F{pqr}{}\cdot{}{}\cdot{}s and \R pqs\cdot\cdot\ in
$\varphi(si;pqr)$ and $R(si;pq)$ are unitary matrices in the
path basis which do not depend on the source flags
$i=1,2,\ldots,n_s$; they precisely coincide with the matrices
introduced in \erf{coc} and \erf{coa} above.

It is straightforward to derive the polynomial equations \erf{penf}
and \erf{hexf} for the intertwiners \coa\ and \coc\ by using the path
calculus. Namely, since the terms in these identities correspond
to special intertwiners among path Hilbert spaces,
they can be easily given in the corresponding path basis; for example,
$(\cop\oti\id\oti\id)(\coa)\colon\;\HS(11,12,21,22)\to\HS(111,112,12,2)$
reads
  $$   (\cop\oti\id\oti\id)(\coa) =
  \bigoplus_{t,i}\bigoplus_{p,q,r,s}\, \sum_{u,v,w}
  \sum_{\alpha,\beta,\atop\gamma,\delta} \sum_{\mu}
  \vert \PL(ti;\beta_{vs}^t,\alpha_{ur}^v,
  \mu_{pq}^u)\rangle\, \F {urs}\alpha v\beta\gamma w\delta t
  \, \langle \PR(ti;\delta_{uw}^t, \gamma_{rs}^{w},\mu_{pq}^u)\vert \,.
  $$ \vskip -8.2mm \be {} \ee
Note that the present construction does not involve the \CGC s at all. Thus
in particular in the path construction it is manifest that the matrices $R$
and $F$ are completely independent of the values of the dimensionalities
$n_r$ of the simple ideals of \h.

%

Since the $R$ and $F$ coefficients can be described in the path basis,
they are almost independent of the ${\cal U}_2$ gauge choice. Namely, a
gauge transformation of \coc\ and \coa\ transforms
simultaneously the path basis through the transformation of the
rank one projectors in $(e_p\otim e_q)\cdot\cop(\e rii)$; as a
consequence, the numerical coefficients $R$ and $F$ are almost invariant.
However, certain phase choices remain undetermined, because the path basis
cannot fix the phases of the image vectors of the aforementioned rank
one projections, since they determine only rays of unit vectors.
Therefore one can use a transformation $U\in{\cal U}_2$ of the form
  \be  U=U_0\oplus \left( \UN\otim\UN-\cop(\UN)\right) \,, \ee
with
  \be  U_0=\bigoplus_{p,q,r\in\hat H}\,\bigoplus_{i=1}^{n_r}
  \bigoplus_{\alpha=1}^{N_{pq}^{\ \,r}}
  \vert{\bf p}(ri;\alpha_{pq}^r)\rangle\,\zeta_\alpha^{(pq)_{\scriptstyle r}}\,
  \langle{\bf p}(ri;\alpha_{pq}^r)\vert \,, \ee
where $\zeta_\alpha^{(pq)_{\scriptstyle r}}$ are pure phases, to change
some of the phases of the $F$ coefficients:
  \be  (F_U^{})^{(pqr)_{\scriptstyle t}}_{\alpha u\beta,\gamma v\delta}=
  \zeta_\alpha^{(pq)_{\scriptstyle u}}\zeta_\beta^{(ur)_{\scriptstyle t}}\cdot
  \Fpqr \cdot \bar\zeta_\delta^{\,(pv)_{\scriptstyle t}}
  \bar\zeta_\gamma^{\,(qr)_{\scriptstyle v}} \,.  \labl{uu}
Choosing $\zeta^{(pq)_{\scriptstyle s}}=1$ whenever $p$ or $q$ equals zero,
we keep our phase choice that $\Fpqr =1$ whenever $p,\,q,$ or
$r$ is the trivial sector and $F$ is nonvanishing.
We can then employ the residual gauge freedom to fix the phases of other
$F$-coefficients. For example, with $\zeta^{(p\bar p)_{\scriptstyle0}}$
and $\zeta^{(\bar pp)_{\scriptstyle0}}$ one can fix
the components \FF p\bp p00p to positive values in the case
of non-selfconjugate sectors $p$. (In the case of selfconjugate sectors
these coefficients are real due to the square identities \erf{sqi}, and the
sign which indicates the reality or pseudoreality of the sector cannot
be changed.)

\subsection{Labelled tetrahedra}

In a path $\vert{\bf p}(\cdo)\rangle$ the $\hh$-label lines in a
pyramid are oriented upwards, while in the dual paths $\langle{\bf
p}(\cdo)\vert$ they are oriented downwards.
According to \erf{33}, \Fpqr\ thus corresponds to the transition
  \be
  \bP(99,70) \mpliv00{20}{20}211{20}{13} \puliv{40}0{-1}1{20}{13}
  \puliv{80}0{-1}1{40}{13} \puliv{40}{40}01{30}{20}
  \putsm{2.9}{10.5}p \putsm{33.2}{10.5}q \putsm{73.2}{10.5}r
  \putsm{22.7}{30.5}u \putsm{43.2}{54.4}t
  \putsc{14.5}{20}\alpha \putsc{42}{41}\beta \eP
            \raisebox{3.3 em}{$\longleftarrow$}\hsp{2.1}
  \bP(90,70) \mpliv{40}{40}{20}{-20}21{-1}{20}{13}
  \puliv{60}{20}{-1}{-1}{20}{13} \puliv{40}{40}{-1}{-1}{40}{33}
  \puliv{40}{70}0{-1}{30}{16}
  \putsm{2.9}{10.5}p \putsm{42.9}{10.5}q \putsm{73.2}{10.5}r
  \putsm{53.0}{30.5}v \putsm{43.2}{54.4}t
  \putsc{62.4}{20}\gamma \putsc{34}{40.8}\delta \eP
  \labl g
By glueing the path and dual path together,
one can therefore represent \Fpqr\ as a tetrahedron
with oriented edges labelled by $p,q,r,u,v,t$ and with vertices labelled by
$\alpha,\beta,\gamma,\delta$, with the order of the matrix
indices $\alpha u\beta,\gamma v\delta$ of \F{pqr}{}\cdot{}{}\cdot{}s
compatible with the prescribed orientation of $u$ and $v$:
  \be  \raisebox{2 em}{\Fpqr$\heq$} \qquad
  \tetrA pqruvt\gamma\alpha{\delta \mbox{\hspace{22.2mm}} \beta}
  \labl{pic2} \vskip 4.2mm
This graphical description indicates that the matrices $F$ correspond to
the $6j$-symbols of the recoupling theory of groups \cite g and
quantum groups \cite{qg}, to a (partially gauge-fixed) version
of the fusing matrices
 \futnote{This correspondence is already implicit in \cite{drin}.
Note, however, that in \cft\ usually `gauges' are considered
in which the fusing matrices are not unitary.}
 {\sf F} \cite{mose,liyu} of \cft,
and also to the Boltzmann weights for triangulations of three-manifolds
\cite{tuvi,d,bawe} giving rise to topological lattice field theories
\cite{d,chfs}. Explicitly, the relation with the fusing matrices reads
  \be  \Fpqr\heq{\sf F}_{uv} \mbox{{\Large[}$\begin{array}{cc} {}\\[-1.55em]
  \!\!{\scs p}&\!\!\!\!{\scs q}\!\! \\[-.32em] \!\!{\scs t}&\!\!\!\!{\scs r}
  \!\! \end{array} {\mbox{\Large]}}_{\alpha\beta}^{\gamma\delta}$} \;, \ee
while the relation with $6j$-symbols is (in the case where all
fusion coefficients are 0 or 1)
  \be  \FF pqruvt\heq \mbox{ {\Large\{}$\begin{array}{ccc} {}\\[-1.55em]
  \!\!{\scs p}&\!\!\!\!{\scs q}&\!\!\!\!{\scs u}\!\!\!\! \\[-.32em]
  \!\!{\scs r}&\!\!\!\!{\scs t}&\!\!\!\!{\scs v}\!\!\!\!
  \end{array}$ {\Large\}} } \!. \ee

 The geometrical contents of the pentagon equation \erf{penf}
can be best understood by using in place of the graphical representation
\erf{pic2} the diagram that is dual to it in the sense of \twodim\
simplicial complexes, i.e.\ where an edge is replaced by a transversal
edge while a trivalent vertex is replaced by a triangle.
The dual diagram of a tetrahedron is again a tetrahedron,
but the advantage is that, as will be shown below, now the pentagon
equation acquires a transparent meaning as a move in triangulations of
three-manifolds. The edges of
the dual tetrahedron are still labelled by the sector indices $p,q,...\,$,
but the multiplicity indices $\alpha,\beta,...$ are now associated to
the faces rather than to the vertices of the tetrahedron, according to
  \be  \raisebox{2 em}{\Fpqr$\heq$} \qquad
  \tetraf pqruvt\alpha\beta {\dbx\delta \mbox{\hspace{11mm}} \dbx\gamma}
  \labl{pic3} \vskip 4.2mm\noindent
Here the faces whose labels are framed by a solid line are in positive
orientation, and those where the framing is by a dashed line are in
negative orientation.
The representation \erf{pic3} corresponds to representing paths by
collections of triangles that are glued together along their common edges
rather than by pyramids, with the indices $\alpha$ etc.\
labelling the faces; e.g.\ in place of \erf g one writes
  \be
  \bP(80,70) \mpliv0{30}{30}{30}21{-1}{30}{18} \puliv0{30}11{30}{18}
  \puliv{60}{30}{-1}{-1}{30}{18} \puliv0{30}10{60}{33}
  \putsm{8.2}{45.7}p \putsm{47.9}{45.7}q \putsm{47.3}{8.5}r
  \putsm{8.2}{8.5}t \putsm{40.1}{32.5}u
  \putsc{24.6}{38.1}{\fbx\alpha} \putsc{24.6}{12.9}{\fbx\beta} \eP
     \hsp1 \raisebox{2.6 em}{$\longleftarrow$}\hsp{2.1}
  \bP(80,70) \mpliv0{30}{30}{30}21{-1}{30}{18} \puliv0{30}11{30}{18}
  \puliv{60}{30}{-1}{-1}{30}{18} \puliv{30}{60}0{-1}{60}{33}
  \putsm{8.2}{45.7}p \putsm{47.9}{45.7}q \putsm{47.3}{8.5}r
  \putsm{8.2}{8.5}t \putsm{32.5}{12.5}v
  \putsc{37.8}{24.8}{\fbx\gamma} \putsc{12.6}{24.8}{\fbx\delta} \eP
  \ee

\subsection{Invariants of three-manifolds}

In the description of tetrahedra according to \erf{pic3}, the two sides
of the pentagon equation \erf{penf} correspond to two
alternative slicings of a pentahedron into tetrahedra: either one
inserts an additional face into the pentahedron such as to
obtain two tetrahedra glued together along this face, or one inserts
an additional edge (linking the vertices that do not belong to the
face used previously) such as to obtain three tetrahedra sharing this
particular edge.
 \futnote{In the theory of triangulations of three-manifolds, this move
goes under various names, e.g.\ as Matveev move \cite{tuvi}, as the
(2,3)-move \cite{move1}, or as the triangle-link exchange \cite{move2}.}
 The summations in \erf{penf} are then precisely over those faces and/or
non-boundary links that are shared by different tetrahedra.
Thus, pictorially the pentagon identity may be written as
  \be  \begin{array}{l} \hsp1
  \tetraf pqvuyt\alpha\sigma{\dbx\mu    \mbox{\hspace{10.3mm}} \dbx\beta}
  \tetraf ursxvt\delta\nu   {\dbx\sigma \mbox{\hspace{10.3mm}} \dbx\gamma}
  \\ {} \\[.6 mm] \hsp5 \raisebox{2 em}{$={\displaystyle\sumh w}$} \qquad
  \tetraf pqruwx\alpha\delta{\dbx\lambda\mbox{\hspace{10.3mm}} \dbx\eta}
  \tetraf pwsxyt\lambda\nu  {\dbx\mu    \mbox{\hspace{10.3mm}} \dbx\kappa}
  \tetraf qrswvy\eta\kappa  {\dbx\beta  \mbox{\hspace{10.3mm}} \dbx\gamma}
  \end{array}\ee \vskip3.8mm \noindent
where the summation over all faces that appear twice (once with
positive and once with negative orientation) is implicit.

Similarly, using the pictorial representation
  \be  \raisebox{2 em}{\Fbpqr$\heq$} \qquad
  \tetrab pqruvt\alpha\beta {\fbx\delta \mbox{\hspace{11mm}} \fbx\gamma}
  \labl{pic4} \vskip 2mm\noindent
the unitarity property \erf u is expressed by
  \be  \begin{array}{l}
  \raisebox{2 em}{${\displaystyle\sumh w}$} \qquad
  \tetraf pqruwt\alpha\beta {\dbx\mu    \mbox{\hspace{10.3mm}} \dbx\nu}
  \tetrab pqrvwt\gamma\delta{\fbx\mu    \mbox{\hspace{10.3mm}} \fbx\nu}
     \hsp{-1} \raisebox{2 em}{$=\delta_{\alpha\gamma}\delta_{\beta\delta}
           \delta_{uv}\,.$} \qquad
  \end{array}\ee \vskip3.8mm \noindent

By combining the unitarity with the pentagon identity, one arrives
at another equation which has an interesting graphical interpretation.
Namely, consider the pentagon identity in the form
  $\sum_\eta \F{wzv}\nu p\eta\sigMa y\mu t \F{pqr}\alpha u\beta\gamma v\eta t
  =\sum_x \sum_{\kappa,\lambda,\rho} \F{wzq}\nu p\alpha\rho x\kappa u
  \F{wxr}\kappa u\beta\lambda y\mu t \F{zqr}\rho x\lambda\gamma v\sigMa y$,
and multiply it with
  \be  \sumh{w,y,z}\, \sumn\mu wyt \sumn\nu wzp \sumn\sigMa zvy
  d_w d_z\, \Fb{wzv}\nu p\delta\sigMa y\mu t \,. \ee
Then on the left hand side, by the use of unitarity we obtain
  \be  \sum_{w,y,z}d_w d_z \sum_{\sigMa,\mu,\nu}
  \F{wzv}\nu p\eta\sigMa y\mu t \Fb{wzv}\nu p\delta\sigMa y\mu t
  =\sum_{w,z}d_w d_z\,N_{wz}^{\ \;p}\,\delta_{\eta\delta}
  =\delta_{\eta\delta}\, d_p\,\sumh w d_w^2 \,,  \ee
where in the last equality we made use of the fact that
$\sumh r\npqr\,d_r=d_p d_q$ and that $\npqr=N_{p\bar r}^{\ \,\bar q}$
and $d_{\bar r}=d_r$.
Also note that according to \erf{s'}, $\sum_w d_w^2=\sigmASQ$ is the
order of \h. Thus we arrive at the relation
  \be  d_p\,\sigmASQ\, \Fpqr=\sum_{w,x,y,z}\sum_{\kappa,\lambda,\mu,
  \atop\nu,\rho,\sigMa} d_w d_z\, \F{wzq}\nu p\alpha\rho x\kappa u
  \F{wxr}\kappa u\beta\lambda y\mu t \F{zqr}\rho x\lambda\gamma v\sigMa y
  \Fb{wzv}\nu p\delta\sigMa y\mu t \,. \labl S
In graphical notation, this means
  \be  \begin{array}{l} \hsp3
  \tetraf pqruvt\alpha\beta {\dbx\delta \mbox{\hspace{11mm}} \dbx\gamma}
  \\ {} \\[-.6 mm] \raisebox{2 em}{$={\displaystyle\sumh{w,x,y,z}}$}\qquad
  \tetraf wzqpxu\nu\alpha   {\dbx\rho   \mbox{\hspace{10.3mm}} \dbx\kappa}
  \tetraf wxruyt\kappa\beta {\dbx\lambda\mbox{\hspace{10.3mm}} \dbx\mu}
  \tetraf zqrxvy\rho\lambda {\dbx\gamma \mbox{\hspace{10.3mm}} \dbx\sigMa}
  \tetrab wzvpyt\nu\delta   {\fbx\sigMa \mbox{\hspace{10.3mm}} \fbx\mu}
  \\{} \end{array}\labl{pic5} \vskip2.4mm \noindent
where we did not care about the explicit factors of $d_s$
in \erf S. To understand the latter factors, it is helpful to rewrite
our formul\ae\ in terms of the normalized quantity
  \be  \Fhpqr:=\Fpqr/\sqrt{d_u d_v} \,.  \ee
Then the pentagon identity becomes
  \be  \sumn\sigma uvt \Fha{pqv}\sigma u\alpha \mu y\beta  t
                 \cdot \Fha{urs}\nu x\delta \sigma v\gamma t
  = \sumh w d_w \sumn\kappa wsy \sumn\lambda pwx \sumn\eta qrw
                       \Fha{pqr}\delta u\alpha \lambda w\eta x
                 \cdot \Fha{pws}\nu x\lambda \mu y\kappa t
                 \cdot \Fha{qrs}\kappa w \eta \beta v\gamma y \,, \labl E
while \erf S reads
  \be  \sigmASQ\, \Fhpqr=\sumh{w,x,y,z}\hsp{-.8} d_w d_x d_y d_z\,
  \sumn\kappa wxu \sumn\lambda xry \sumn\mu wyt \sumn\nu wzp \sumn\rho zqx
  \sumn\sigMa zvy \Fh{wzq}\nu p\alpha\rho x\kappa u
  \Fh{wxr}\kappa u\beta\lambda y\mu t \Fh{zqr}\rho x\lambda\gamma v\sigMa y
  \Fhb{wzv}\nu p\delta\sigMa y\mu t \,. \labl F \vskip1.8mm \noindent
The geometrical contents of this equation can be read off the picture
\erf{pic5}. Namely, \erf F describes the subdivision of a tetrahedron into
four tetrahedra that is obtained by adding a further vertex which is
joined to the four original vertices by the edges labelled $w,x,y,z$.
The summation is then over these edges as well as over the six faces
that are shared by any two of the new tetrahedra.  Note that in \erf E
and \erf F each internal edge is accompanied by a factor $d_r$. Similarly
the additional factor of $\sigmASQ$ on the left hand side of \erf F
can be accounted for by attaching a factor of $\sigmASQI$ to each vertex.
The latter amounts to a sort of renormalization that takes place
in the transition to a finer lattice.

Consider now a three-dimensional compact oriented manifold ${\cal M}$
(possibly with boundary $\partial{\cal M}$) and a triangulation
$\cal T$ of $\cal M$ (with induced triangulation $\partial{\cal T}$
on $\partial{\cal M}$) by tetrahedra $T$ with faces $F$ and edges $L$. Further,
define the quantity
  \be Z({\cal M},{\cal T}):={\sigmASQ}^{|{\cal T}|-|\partial{\cal T}|/2}_{}
  \sum_{\{p_L^{},\alpha_F^{}\}}\,
  \prod_{L\in{\cal T}\setminus\partial{\cal T}}\!\! d(L) \!
  \prod_{T\in{\cal T}\setminus\partial{\cal T}}\!\! \hat F(T) \,. \ee
Here $|{\cal T}|$ and $|\partial{\cal T}|$ denote the number of
vertices of $\cal T$ and $\partial{\cal T}$, respectively;
the sum is over all possible assignments $L\mapsto p_L^{}\in\hh$
of sector labels to all edges $L$ of $\cal T$ and $F\mapsto \alpha_F^{}$
of multiplicity labels to all faces $F$ of $\cal T$; finally
$d(L)=d_{p_L^{}}^{}$ and $(\hat F(T))^{(pqr)_{\scriptstyle t}}_
{\alpha u\beta,\gamma v\delta}=\Fh{p_{L}^{}q_{L}^{}r_{L}^{}}{\alpha_{F}^{}}
{u_{L}^{}}{\beta_{F}^{}}{\gamma_{F}^{}}{v_{L}^{}}{\delta_{F}^{}}{t_{L}^{}}$
with the appropriate
choices of the various edges and faces as prescribed by the tetrahedron $T$.

It can be shown \cite{tuvi,d,chfs} that invariance of $Z$ under the
pentagon move and under subdivision (in the sense of the validity of
\erf E and \erf F) implies that for sufficiently fine triangulations
the number $Z({\cal M},{\cal T})$ is in fact independent of $\cal T$
and hence defines a topological invariant of the three-manifold $\cal M$.
In other words, $Z$ is invariant under
smooth changes of the metric of the manifold; thus it can be interpreted
as the partition function of a topological field theory \cite{tuvi,d,chfs}.

\sect{Classification of \rha s with $\vert\hh\vert\leq3$}

As mentioned in the introduction, among the fusion rule algebras
${\cal M}$, only the modular ones are relevant to the classification of
\rha s \h\ with $\vert \hh\vert={\rm dim}\, {\cal M}$.
Thus we need to consider only such coproducts
that provide us modular fusion rules. Modular fusion rules
have been completely classified
up to ${\rm dim}\, {\cal M}\leq 3$: in addition to the trivial \onedim\
\furu\ \alg, there are two \furu\ \alg s with ${\rm dim}\, {\cal M}=2$,
and three with ${\rm dim}\, {\cal M}=3$ \cite{capo2}.
In the subsections 4.1 to 4.5 below, we will classify the
\rha s satisfying these fusion rules. Apart from the classification of
$\vert\hh\vert\leq 3$ \rha s we also describe, in section 5,
the $\vert\hh\vert=4$ \rha s corresponding to the fusion rules of the
level 3 integrable representations of the \Ao\ affine Lie algebra.
We note that conjugate pairs of solutions always exist,
corresponding to the conjugate representation of the modular group $\Gamma$
which is generated by the complex conjugate matrices of the generators
$S$ and $T$. The corresponding
conjugate models possess complex conjugate statistical phases, while their
central charges are related by $c+c'=0\;{\rm mod}\,8$.

As already mentioned (see also section \ref{secnp} below), the
dimensionalities $n_r$ are not fixed by the properties of \rha s.
In the explicit description of the coproduct matrices and
statistics operators we will use the
minimal (integer) multiplicities $n_r$ that are allowed by the fusion
rules (i.e.\ satisfy \erf{npqr}). As it turns out, in all the examples
considered so far these are determined via
  \be  d_r\leq n_r<d_r+1  \ee
by the statistical dimensions $d_r$ of the sectors.

It also turns out that typically some of the parameters that
appear in the solutions to the polynomial equations
are absent from the results for the left inverses and the monodromy matrix.
These parameters are therefore irrelevant and we expect
 \futnote{We plan to discuss this issue in more detail elsewhere.}
that they can be fixed to 1 by ${\cal U}_2$ transformations of the
type \erf{uu} (this was explicitly checked in the
Ising case \cite V). Therefore in our classification we count only
those parameters that appear in the statistics parameters and in
the monodromy matrix.

\subsection{The $\zet_2$ fusion rules}

\noindent The fusion rule matrices $N_p$ (i.e., the matrices
with entries $(N_p)_{qr}=\npqr$) defining the $\zet_2$ \furu s are
  \be  N_0=\left(\matrix{1&0\cr 0&1\cr}\right),\qquad
  N_1=\left(\matrix{0&1\cr 1&0\cr}\right). \labl{z2}
The corresponding coproduct matrix $M_\Delta$ reads
  \be  \copmatab{1\\0}  \raisebox{-1.1em}. \labl{c2}
By solving the pentagon and hexagon equations, we find that the
non-trivial intertwiner data read
  \be  \FF111001 =\omega^2,\qquad \RR110=\omega, \qquad
   r=e_0+e_1,\qquad l=e_0+\omega^2 e_1,  \ee
with $\omega$ an arbitrary fourth root of unity,
  \be  \omega^4=1 \,. \labl{44}
Further, the statistics operators $\stopt_{00},\,\stopt_{01}$
and $\stopt_{10}$ are all equal to $\UN$, while
the single non-trivial statistics operator is found to be
  \be  \stopt_{11}=\omega\cdot\UN \,. \ee
The left inverse $\Phi_0$ is, as always, equal to $\id_H$, while
the non-trivial left inverse reads
  \be  \Phi_1(f)=\left\{\begin{array}{ll} e_1,& f=e_0,\\[.4 mm]
                            e_0,& f=e_1.\end{array}\right. \ee
The statistics parameters are $\lambda_0=\UN$ and $\lambda_1=\omega\UN$,
and for the monodromy matrix we find $Y=y\,\UN$ with
  \be  y=\left(\matrix{1&1\cr 1&\omega^2\cr}\right). \ee

According to \erf{44}, there are four inequivalent candidates for
\rha s which possess the $\zet_2$ \furu s; they are parametrized by the
fourth root of unity which is chosen for the parameter $\omega$. In the
table below we list these \alg s. The second and third columns
provide the statistics phases $\omega_0=\exp(2\pi\ii w_0)$ and
$\omega_1=\exp(2\pi\ii w_1)$ of the sectors (the
$w_r=(1/2\pi\ii)\cdot\log\omega_r$ are chosen such that they lie in
the interval [0,1)\,);
the next column contains the number $|\sigmA|^2$ defined according
to \erf{sig}; if the algebra is not degenerate (in the sense
described after \erf c), then $c\;$mod\,8 is as listed in the fifth
column; finally, in the last column we indicate whether the \alg\ is
degenerate, and if not, present known examples of \cfts\ which
match the required values of the conformal weights $\omega_p$
modulo integers and of $c$ modulo 8.
  \be \begin{tabular}{r|cccc|r}
  \multicolumn{1}{c|}{$\omega$} & $w_0$ & $w_1$ & $|\sigmA|^2$ & $c$
  & \interpr \\[.7 mm] \hline &&&&&\\[-2.7 mm]
  \ii   & 0 & 1/4 &  2 & 1 & \hz A11~ \\[1.3 mm]
  $-1$  & 0 & 1/2 &  0 & . & \degen~  \\[1.3 mm]
  $-\ii$& 0 & 3/4 &  2 & 7 & \hz E71~ \\[1.3 mm]
  1     & 0 &   0 &  4 & . & \degen~
  \end{tabular} \ee
Here \wz A11 stands for (one chiral half of) the WZW theory based
on the affine Lie algebra \Ao\ at level 1, etc.

\subsection{The \LY fusion rules}

\noindent The fusion rules defined by the matrices
  \be  N_0=\left(\matrix{1&0\cr 0&1\cr}\right),\qquad
  N_1=\left(\matrix{0&1\cr 1&1\cr}\right) \labl{ly}
are known as the \LY fusion rules, because they are the
\furu s of the $(p,q)=(5,2)$ non-unitary minimal
conformal model of the Virasoro \alg\ which describes the \LY edge
singularity. These \furu s
will lead to an example for a \RHA\ that provides the third smallest index
$(3+\sqrt{5})/2$ in the Jones \cite{J,ind} classification.\\
The coproduct matrix for \erf{ly} is
  \be  \copmatabb {1_1\\0\\1_1} {1_2\\ \\1_2} \raisebox{-1.7em}. \labl{cy}
The non-trivial intertwiner data are found to be
  \be  \begin{array}{l}  \FF111110=1, \quad \FF111001=-\FF111111=z^2, \\[2.4
mm]
  \FF111011=\FFb111101=\omeG  z, \quad \RR110=\zeT ^2, \quad
  \RR111=\zeT , \\[2.4 mm]
  r=l=e_0+\left({1+\sqrt5\over 2}\right)^{1/2}\cdot \e111,
  \end{array}\ee
where
  \be  z=\sqrt{\frac12\,(\sqrt5-1)} \labl{513}
and
  \be  |\omeG | = 1  \,, \qquad  \zeT =\exp(\pm3\pi\ii/5)
  ={1\over 4}\cdot\left(1-\sqrt5\pm\ii\sqrt{10+2\sqrt{5}}\,\right). \labl{514}
Because of the relation \erf{nu0}, the statistics operators
$\stopt_{pq}=\stopt(\mu_p;\mu_q)$ for $p=0$ or $q=0$ are immediate;
for the single remaining non-trivial statistics operator we obtain
  \be
  \stopt_{11}^{}=\left( \begin{array}{cccc}
  \zeT^2 e_0^{}+\zeT\e122-z^2\zeT^3\e111 & 0 & -\omeG z\zeT^4\e112 & 0
  \\[1.2 mm]   0 & -z^2\zeT^3\e111 & 0 & -\omeG z\zeT^4\e112
  \\[1.2 mm] -\bmeG z\zeT^4\e121 & 0 & \zeT e_0^{}-z\e122 & 0 \\[1.2 mm]
  0 & -\bmeG z\zeT^4\e121 & 0 & \zeT e_0^{}-z\e122 \end{array} \right) .  \ee
For the non-trivial left inverse we find
  \be \Phi_1([f])=\left\{\begin{array}{ll}
     z^2\,\e1ij,& [f]={[e_0^{}]}^{i,j},\\[1 mm] z\,\e1ij,&
  [f]={[\e122]}^{i,j}, \\[1 mm] e_0,& [f]={[\e111]}^{1,1},\\[1 mm]
     0,& {\rm otherwise.} \end{array}\right. \ee
 Hence the statistics parameters are
  \be  \lambda_0=\UN, \qquad \lambda_1=-\zeT^3z^2\,\UN \labl{517}
so that $d_1=z^{-2}=(1+\sqrt{5})/2$. The monodromy matrix reads
  \be y=\left(\matrix{1&{1+\sqrt{5}\over 2}\cr
  {1+\sqrt{5}\over 2}&-1\cr}\right). \labl{518}
The possible \RHA s are then classified by the two possible values of
the parameter $\zeT $. We obtain
  \be \begin{tabular}{r|cccc|r}
  \multicolumn{1}{c|}{$\zeT $} & $w_0$ & $w_1$ & $|\sigmA|^2$ & $c$
  & \interpr \\[.7 mm] \hline &&&&&\\[-2.7 mm]
  $\exp(2\pi\ii{3\over 10})$
   & 0 & $\frac25$ & ${5+\sqrt{5}\over 2}$ & $\frac{14}5$ & \hz G21~ \\[1.3 mm]
  $\exp(2\pi\ii{7\over 10})$
   & 0 & $\frac35$ & ${5+\sqrt{5}\over 2}$ & $\frac{26}5$ & \hz F41~
  \end{tabular} \ee
Note that this time no degenerate solutions exist.

\subsection{The $\zet_3$ fusion rules}

\noindent The $\zet_3$ fusion rules are defined by the following
fusion rule matrices:
  \be  N_0=\left(\matrix{1&0&0\cr 0&1&0\cr 0&0&1\cr}\right),\qquad
  N_1=\left(\matrix{0&1&0\cr 0&0&1\cr 1&0&0\cr}\right),\qquad
  N_2=\left(\matrix{0&0&1\cr 1&0&0\cr 0&1&0\cr}\right).  \labl{z3}
The corresponding coproduct matrix is
  \be  \copmatabc {1\\2\\0} {2\\0\\1} \raisebox{-1.7em}. \ee
We find the following non-trivial intertwiner data:
  \be  \begin{array}{l} \FF121001=\FFb212002=\bmega\FF222110=
  \omega \FF111220=\zeta_1,\\[2.6 mm]
  \FF112201=\bmega\FFb122012=\zeta_2, \qquad
  \FF221102=\bmega\FFb211021=\zeta_1\zeta_2, \qquad  \\[2.6 mm]
  \RR112=\RR221=\omega,\qquad \RR120=\omega^2\bzeta_1,\qquad
  \RR210=\omega^2\zeta_1,\\[2.6 mm]
  r=e_0+e_1+e_2,\qquad l=e_0+\zeta_1e_1+\bzeta_1e_2 \,. \end{array}\ee
The three parameters $\omega,\;\zeta_1,\;\zeta_2$  are restricted by
  \be  \omega^3=1,\quad \vert\zeta_1\vert= \vert\zeta_2\vert=1. \ee
The non-trivial statistics operators are
  \be  \begin{array}{l} \stopt_{11}=\stopt_{22} =\omega\UN,\qquad
  \stopt_{12}=\omega^2\bzeta_1
  (e_0^{}+\zeta_1\bzeta_2e_1^{}+\zeta_1^2\zeta_2^{} e_2^{}),\\[2.4 mm]
  \stopt_{21}=\omega^2\zeta_1 (e_0+\bzeta_1\zeta_2 e_1+
  \bzeta_1^2\bzeta_2 e_2)=\omega\stopt_{12}^*, \end{array}\ee
and the non-trivial left inverses read
  \be  \Phi_1(f)=\left\{\begin{array}{ll}
       e_1,\quad f=e_0,\\[.4 mm] e_2,\quad f=e_1,\\[.4 mm]
       e_0,\quad f=e_2. \end{array}\right.\qquad\quad
  \Phi_2(f)=\left\{\begin{array}{ll}
       e_2,\quad f=e_0,\\[.4 mm] e_0,\quad f=e_1,\\[.4 mm]
       e_1,\quad f=e_2. \end{array}\right. \ee
The non-trivial statistics parameters are $\lambda_p=\omega\,\UN$ for
$p=1,2$, and for the monodromy matrix we obtain
  \be  y=\left(\matrix{1&1&1\cr 1&\omega^2&\omega\cr
    1&\omega&\omega^2\cr}\right). \ee
Note that the left inverses and monodromy matrix are independent of
the parameters $\zeta_1$ and $\zeta_2$.
As a consequence, the candidates for the \RHA s are characterized by
the three possible values of the parameter $\omega$ alone:
  \be \begin{tabular}{c|ccccc|r}
  \multicolumn{1}{c|}{$\omega$} & $w_0$ & $w_1$ & $w_2$ & $|\sigmA|^2$ & $c$
  & \interpr \\[.7 mm] \hline &&&&&\\[-2.7 mm]
  $\exp(2\pi\ii/3)$ & 0 & 1/3 & 1/3 & 3 & 2 & \hz A21~ \\[1.3 mm]
  $\exp(4\pi\ii/3)$ & 0 & 2/3 & 2/3 & 3 & 6 & \hz E61~ \\[1.3 mm]
  1                 & 0 &  0  &  0  & 9 & . & \degen~
  \end{tabular} \ee

\subsection{The Ising fusion rules}

\noindent The fusion rules defined by the matrices
  \be  N_0=\left(\matrix{1&0&0\cr 0&1&0\cr 0&0&1\cr}\right),\qquad
  N_1=\left(\matrix{0&1&0\cr 1&0&1\cr 0&1&0\cr}\right),\qquad
  N_2=\left(\matrix{0&0&1\cr 0&1&0\cr 1&0&0\cr}\right)  \labl{is}
are known as the Ising \furu s. The corresponding
\rha s have already been described in \cite{V}. For
completeness we present the table of the solutions and of
the associated \cft\ models. There are eight inequivalent \rha s,
parametrized by a number $\omega$ which must be an odd 16th root of
unity and coincides with the statistics phase of the non-abelian sector.
For any choice of this root, one has a realization in terms
of the \wZ Br1 \kma; in addition, for some specific choices
of $\omega$ there exist other \cfts\ sharing the relevant \rha.
  \be \begin{tabular}{c|ccccc|r}
  \multicolumn{1}{c|}{$\omega$} & $w_0$ & $w_1$ & $w_2$ & $|\sigmA|^2$ & $c$
  & \interpr \\[.7 mm] \hline &&&&&\\[-2.7 mm]
  $\exp((2r+1)\pi\ii/8)$ & 0 & 1/2 & $(2r+1)/16$ & 2 & $(2r+1)/2$ &
                          \hz Br1,\ $r>1$ \\[1.3 mm]
  $\exp( \pi\ii/8)$ & 0 & 1/2 & 1/16 & 2 & 1/2 &
                          \mbox{\small Ising model} \\[1.3 mm]
  $\exp(3\pi\ii/8)$ & 0 & 1/2 & 3/16 & 2 & 3/2 & \hz A12 \\[1.3 mm]
  $\exp(15\pi\ii/8)$& 0 & 1/2 &15/16 & 2 &15/2 & \hz E82
  \end{tabular} \labl{iscl}

\subsection{The `(7,2)' fusion rules}

\noindent Consider the following fusion matrices:
  \be  N_0=\left(\matrix{1&0&0\cr 0&1&0\cr 0&0&1\cr}\right),\qquad
  N_1=\left(\matrix{0&1&0\cr 1&0&1\cr 0&1&1\cr}\right),\qquad
  N_2=\left(\matrix{0&0&1\cr 0&1&1\cr 1&1&1\cr}\right).  \labl{72}
These describe the fusion rules of the $(p,q)=(7,2)$ non-unitary minimal
conformal model of the Virasoro \alg, and hence we refer to them as
the `(7,2)' fusion rules. The coproduct matrix for \erf{72} reads
  \be  \copmatabbccc {1_1\\0\\2_1\\2_1\\2_2\\2_3}
       {1_2\\2_2\\2_3\\1_1\\1_2\\{~} } {2_1\\2_1\\1_1\\0\\1_1\\2_1}
       {2_2\\2_2\\1_2\\1_2\\2_3\\{~} } {2_3\\2_3\\ \\2_2\\{~} \\{~} }
  \raisebox{-2.7em}. \ee

 \def\S{x} \def\Ss{\sqrt x} \def\T{z} \def\Ts{\sqrt z}
 \def\bon{\zetab_1} \def\btw{\zetab_2} \def\bth{\zetab_3} \def\bfi{\zetab_6}
 \def\bse{\omega_2} \def\bei{\zetab_4} \def\bni{\zetab_5}
 \def\btf{\omega_1} \def\btfs{\btf\Bse}
 \def\Bon{\bar\bon} \def\Btw{\bar\btw} \def\Bth{\bar\bth}
 \def\Bfi{\bar\bfi} \def\Bse{\bse} \def\Bei{\bar\bei} \def\Bni{\bar\bni}
 \def\Btf{\btf} \def\Btfs{\btf\bse}
 \def\btfSQ{} \def\BseSQ{} \def\BtfSQ{}
 \def\oon{\omega_1} \def\otw{\omega_3} \def\oth{\omega_4}

By solving the square, pentagon and hexagon equations, we find the
following non-trivial intertwiner data. First, the components of the
matrices $F$ are
  \be  \begin{array}{l}
  \FF222220 = 1    ,  \quad
  \FF111222 = \FF212220 = \Btf    ,  \quad
  \FF121110 = \btfs    ,  \quad
  \FF121221 = \bse    ,  \quad
  \FF112221 = \BtfSQ\bse\Bei  ,  \\[2.4 mm]
  \FF112210 = \Btfs\Bei\Bni    ,  \quad
  \FF122210 = \btf\Bni    ,  \quad
  \FF211120 = \bei\bni    ,  \quad
  \FF211221 = \bei    ,  \quad
  \FF221120 = \bni  ,  \\[2.4 mm]

  \FF111001 = \S     ,  \quad
  \FF111221 = -\Btfs\S  ,  \\[2.4 mm]

  \FF111021 = \Btfs\bon^2\btw\bei\bni\Ts      ,  \quad
  \FF111201 = \Bon^2\Btw\Bei\Bni\Ts  ,  \\[2.4 mm]

  \FF112012 = \Btfs\bon^2\btw\bei\bni\S/\Ts     ,  \quad
  \FF112222 = -\Btf\Bon\Bth\Bei\S/\Ts     ,  \quad
  \FF122101 = \Bon^2\Btw\btfSQ\BseSQ\S/\Ts     ,  \\[2.4 mm]
  \FF122221 = -\bse\bon\Bfi\S/\Ts     ,  \quad
  \FF211102 = \Bon^2\Btw\Bei\Bni\S/\Ts  ,  \quad
  \FF211222 = -\bon\Bfi\bei\S/\Ts     ,  \\[2.4 mm]
  \FF221011 = \bon^2\btw\S/\Ts       ,  \quad
  \FF221221 = -\Bon\bfi\S/\Ts      ,  \quad
  \FF222022 = \FFb222202 = \btw\S/\Ts  ,  \\[2.4 mm]

  \FF112022 = -\Btf\btw\Bth\bni\Ss     ,  \quad
  \FF112212 = \FFb122121 = -\Btfs\bon\Ss     ,  \quad
  \FF122201 = -\Btw\bth\Ss  ,  \\[2.4 mm]
  \FF211122 = -\Bon\Ss     ,  \quad
  \FF211202 = -\Btw\Bfi\Bni\Ss     ,  \quad
  \FF221021 = -\btw\bfi\Ss     ,  \quad
  \FF221211 = -\bon\Ss  ,  \\[2.4 mm]

  \FF122112 = \btfSQ\Bse\bei\S/\T     ,  \quad
  \FF122222 = \FF221222 = \FF222221 = -\S/\T  ,  \\[2.4 mm]
  \FF212112 = \bse\S/\T     ,  \quad
  \FF212222 = -\Btf\S/\T     ,  \quad
  \FF221112 = \Bei\S/\T     ,  \quad
  \FF222111 = \btfs\S/\T  ,  \\[2.4 mm]

  \FF122122 = \FF222121 = \btfs\Bon\S^{3/2}/\T   ,  \quad
  \FF122212 = \btf\bon\bei\S^{3/2}/\T   ,  \quad
  \FF212122 = \Bon\S^{3/2}/\T  ,  \\[2.4 mm]
  \FF212212 = \bon\Btfs\S^{3/2}/\T   ,  \quad
  \FF221122 = \Bon\Bei\S^{3/2}/\T   ,  \quad
  \FF221212 = \FF222211 = \bon\S^{3/2}/\T  ,  \\[2.4 mm]
  \FF222012 = -\btw\bfi\S^{3/2}/\T   ,  \quad
  \FF222102 = -\Btw\bth\S^{3/2}/\T  ,  \\[2.4 mm]

  \FF121112 = -\btfs\S^2/\T   ,  \quad
  \FF212111 = -\S^2/\T   ,  \quad
  \FF121222 = \FF222002 = \S^2/\T   ,  \quad
  \FF212221 = \Btf\S^2/\T  ,  \\[2.4 mm]

  \FF121122 = -\Bse\Bon^2\bfi\Bei(\S/\T)^{1/2}   ,  \quad
  \FF121212 = -\bon^2\bth\bei(\S/\T)^{1/2}   ,  \\[2.4 mm]
  \FF212121 = -\Bse\Bon^2\bfi(\S/\T)^{1/2}   ,  \quad
  \FF212211 = -\Btf\bse\bon^2\Bfi(\S/\T)^{1/2}  ,  \\[2.4 mm]

  \FF222112 = \btf\S^3/\T^2 ,   \quad

  \FF222121 = \bth(\S/\T)^{3/2} ,   \quad
  \FF222211 = \bfi(\S/\T)^{3/2} ,   \quad

  \FF222222 = \S^2(\T-1)/\T^2  .
  \\[-1.8 mm] {~} \end{array}\ee
Here $\bon, ...\,,\bni$ are arbitrary phases, $\bfi=\oon \Bth$, and
  \be  \oon^2=1=\bse^2 \,.  \ee
Further, $\T$ is the largest of the three real solutions of the cubic
equation $\T^3+3\T^2-4\T+1=0$, and $\S=2-\T^{-1}$, i.e.
  \be  \begin{array}{l} \T=-2\sqrt{\mbox{$\frac73$}}\,\cos(\mbox{$\frac13$}
  (\pi-{\rm arctan}\,3^{-3/2}))-1 =1-d^{-2} \simeq .692021, \\[2.6 mm]
  \S=d^{-1}\simeq .554958  \end{array} \ee
with
  \be  d:=2\,\cos(\frac\pi7) \,. \ee
For the phases $R$ we find
  \be\begin{array}{l}
  \RR110=\otw^{12}       ,  \qquad    \RR112=\otw^3      ,    \\[2.6 mm]
  \RR121= \RR211=\btfs\otw^9       ,  \qquad
  \RR122= \RR212=\btf\otw^6  ,  \\[2.6 mm]
  \RR220=\otw^2      ,  \qquad\  \RR221=\otw^{10}       ,  \qquad
  \RR222=\otw^9       , \end{array}\ee
where
  \be  \otw=\exp(\pm\ii\pi/7) \,.  \labl{otw}
Finally,
  \be  r= l= e_0^{}+\sqrt d\,\e111+\sqrt{d^2-1}\,\e211 \,. \ee
As is clear from these results, formul\ae\ for statistics operators
and left inverses become rather lengthy in this case, and we refrain from
writing them down.
For the statistics parameters we find
  \be  \lambda_0=\UN \,, \quad \lambda_1=\otw^2\S\,\UN \,, \quad
  \lambda_2=\otw^{10}\S^2\T^{-1}\,\UN \,, \ee
so that in particular
  \be   d_0=1, \quad d_1=d\simeq1.80194, \quad d_2=d^2-1\simeq2.24698. \ee
The monodromy matrix becomes
  \be  y = \left( \begin{array}{ccc} 1 & d & d^2-1 \\[1.9 mm]
  d & \otw^{10}+\otw^6(d^2-1) & \otw^4d+\otw^{10}(d^2-1) \\[1.9 mm]
  d^2-1 & \otw^4d+\otw^{10}(d^2-1) & \otw^8+\otw^{10}d+\otw^4(d^2-1)
  \end{array} \right) . \ee
Using addition theorems for the trigonometric functions as well as the
cubic equation obeyed by $\T$, one sees that all entries of $y$ are
in fact independent of the sign choice in \erf{otw} and real, namely
  \be  y = \left( \begin{array}{ccc} 1 & d & d^2-1 \\[1.1 mm]
  d & 1-d^2 & 1 \\[1.1 mm] d^2-1 & 1 & -d \end{array} \right) . \ee
We thus arrive at the following classification of \rha s
that have the \furu s \erf{72}:
  \be \begin{tabular}{c|ccccc|c} \multicolumn{1}{c|}{$\otw$} &
  $w_0$ & $w_1$ & $w_2$ & $|\sigmA|^2$ & $c$ & \interpr
  \\[.7 mm] \hline &&&&&&\\[-2.7 mm]
  $\exp(\ii\pi/7)$ & 0 & 1/7 & 5/7 & $2d^2+d+1$  & 48/7 & -- \\[1.3 mm]
  $\exp(-\ii\pi/7)$& 0 & 6/7 & 2/7 & $2d^2+d+1$  &  8/7 & $\h[\zet_5^{(0)}]$
  \end{tabular} \labl{tab}
Thus there are two non-degenerate solutions and no degenerate ones.
 \futnote{Note that our results for the central charges $c$ differ
from the ones reported in \cite{capo2} by 4.}
The model denoted by $\zet_5^{(0)}$ is the \cft\ describing the closed
subalgebra generated by the $\zet_5$-neutral primary fields
of the $\zet_5$ parafermion theory (the latter theory which has $c=\frac87$
corresponds to certain self-dual $\zet_5$ critical lattice models \cite{ZF}).
Also, in contrast to all other \rha s considered in this section, at
present no rational field theory that corresponds to the other \rha\ in
table \erf{tab} is known to us. However, the existence of such a solution
is an indication that a corresponding model does exist.

\sect{Rational Hopf \alg s with $\vert\hh\vert>3$}

The calculations leading to the results presented in the
previous section can in principle be performed by hand. However,
for practical purposes, it makes sense to encode the necessary
manipulations into a computer program. We have indeed done so,
 \futnote{In fact, we have written two independent routines, one in MAPLE,
the other in MATHEMATICA, thereby allowing for a convenient cross check
of the results.}
 and it is straightforward to apply the algorithm to any prescribed
\furu\ \alg.
Note that the pentagon and hexagon equations constitute a huge system
of nonlinear equations for a large number of unknowns.
 For instance, in the case of the `(7,2)' \furu s the pentagon
equation provides 748 equations for the 64 $F$s which
are not trivially equal to 1 or 0. That such a system can be
solved by a symbolic manipulation program at all has its origin in the fact
that the system is actually highly over-determined. (E.g. in the
example just mentioned, 319 of the 748 pentagon equations are fulfilled
identically.) To solve the system,
it is convenient to start from the subset of those equations which have
a single monomial on each side and hence upon taking the logarithm
reduce to a {\em linear\/} equation, and separate the phases and
absolute values of these equations. After solving this `linear' part
as well as the constraints provided by the unitarity of the
$F$-matrices, already a lot of the freedom is fixed (in fact, in
all cases considered so far, in particular all absolute values of the
$F$ coefficients are fixed). The remaining nonlinear equations can then be
solved by an iterative procedure where in each step one solves one or
more of the equations `by inspection'
and plugs the result back into the system; while there is no guarantee
that this procedure is fast enough, in all cases we looked at
it actually worked very well.

As an example of a \furu\ \alg\ with ${\rm dim}\, {\cal M}=4$,
let us consider the level 3 affine \Ao\ fusion rules.
The corresponding fusion rule matrices are
  \be  \begin{array}{l}
  N_0=\left(\matrix{1&0&0&0\cr 0&1&0&0\cr 0&0&1&0\cr 0&0&0&1\cr} \right),\qquad
  N_1=\left(\matrix{0&1&0&0\cr 1&0&1&0\cr 0&1&0&1\cr 0&0&1&0\cr}
  \right),\\ {}\\[-1 mm]
  N_2=\left(\matrix{0&0&1&0\cr 0&1&0&1\cr 1&0&1&0\cr 0&1&0&0\cr} \right),\qquad
  N_3=\left(\matrix{0&0&0&1\cr 0&0&1&0\cr 0&1&0&0\cr 1&0&0&0\cr}
  \right). \end{array}\labl{61}
This describes the operator products of the level 3 integrable
representations of the \Ao\ affine Kac\hy Moody algebra.

\noindent The coproduct matrix reads
  \be  M_\Delta=\quad
  \copmatabbccd {1_1\\0\\ \\3\\1_1\\2_1} {1_2\\2_1\\2_2\\ \\1_2\\2_2}
  {2_1\\3\\ \\0\\ \\1_1} {2_2\\1_1\\1_2\\2_1\\2_2\\1_2}
  {3\\2_1\\2_2\\1_1\\1_2\\0} \raisebox{-2.7em}. \labl{62}

It is easy to see that the tensor product $\H\oti \tilde H$ of two
\RHA s \H\ and $\tilde H$, with
  \be  \begin{array}{l}  \cop_{H\otimes\tilde H}(a\otim\tilde a)=
  (a^{(1)}\otim \tilde a^{(1)}) \otim (a^{(2)}\otim \tilde a^{(2)}) \,
  \\[1.5 mm] \cou_{H\otimes\tilde H}(a\otim \tilde a)=\cou(a)\cdot
  \tilde{\cou}(\tilde a) \,,\qquad \apo_{H\otimes\tilde H}(a\otimes \tilde a)=
  \apo(a)\otimes \tilde{\apo}(\tilde a) \,, \end{array} \ee
and with tensor product $\lambda,\rho,R,\varphi, l,r$
quantities, defines again a \RHA. The new fusion matrices are
tensor products of the original fusion matrices,
the statistics parameters are products of the previous
statistics parameters, and the new monodromy matrix is the tensor
product of the previous ones, so that in particular it is invertible.

Now the level 3 affine $A_1^{(1)}$ fusion rules
\erf{61} are actually the tensor product of the
$\zet_2$ \furu s and the \LY \furu s, corresponding to
the following identification of the sector labels:
  \be  \begin{array}{ll}
       0   = 0^{(\zet_2)}_{} \otim 0^{({\rm Lee-Yang})}_{} \,, \quad{} &
       1_i = 1^{(\zet_2)}_{} \otim 1^{({\rm Lee-Yang})}_i  \,, \\[2.1 mm]
       3   = 1^{(\zet_2)}_{} \otim 0^{({\rm Lee-Yang})}_{} \,, &
       2_i = 0^{(\zet_2)}_{} \otim 1^{({\rm Lee-Yang})}_i  \,.
  \end{array}\ee
Therefore the coproduct matrix is within the
${\cal U}_2$-equivalence class of the coproduct matrix that comes
from the tensor product algebra.
 \futnote{Indeed, the matrix \erf{62} is almost identical with the tensor
product of the coproduct matrix \erf{c2} of the $\zets_2$ \furu s and
the matrix for the opposite coproduct \erf{cy} of the \LY \furu s.}
 Hence among the solutions of the pentagon and
hexagon equations we must find the tensor product solutions,
but in principle there can be other solutions as well.

     \def\omegaa{\zeta_1}  \def\bmegaa{\bar\zeta_1}
     \def\omegae{\zeta_2}  \def\bmegae{\bar\zeta_2}
     \def\omegaz{\zeta_3}  \def\bmegaz{\bar\zeta_3}
     \def\zetA{\omega}     \def\zetae{\gamma}
     \def\za{z^2} \def\Za{z^4} \def\ZA{z}
     \def\zb{x} \def\Zb{\bar x}

The pentagon equation now amounts to a total of 752 equations for 71 variables
$F$ which are not trivially 1 or 0 (of these, 375 are fulfilled identically,
and 297 of the remaining equations have on each side only one monomial in
the variables $F$). The most general solution of this system is given by
  \be  \begin{array}{l}
  \FF121110=\FF132210=\FF222220=\FF231120=\FF232113=\FF323112= 1, \\[2.4 mm]
  \FF131223=\FF313221=\FF333003 =\alpha, \qquad
    \FF111223 = \FF131221= \beta, \\[2.4 mm]
  \FF212113=\FF232111= \alpha\beta, \qquad
    \FF133201=\FFb331021 =\omegaa, \qquad
    \FF233102=\FFb332012 =\alpha\omegaa,\\[2.4 mm]
  \FF113023=\FFb311203 =\omegae,\qquad\;
    \FF113221=\FFb311221 =\omegaz,\\[2.4 mm]
  \FF122123=\FF213122=\FFb221213=\FF223013=\FFb223211=\FFb312212
    =\FFb322103=\FF322121=\alpha\omegae,\\[2.4 mm]
  \FFb123112=\FF321112 =\alpha\omegaz,\qquad
    \FF112210=\FFb211120 =\beta\omegaz,\\[2.4 mm]
  \FFb123310=\FFb213320=\FF312230=\FF321130 =\omegaa\omegae,\qquad
  \FFb132212=\FF231122 =\alpha\beta\omegae\omegaz,\\[2.4 mm]
  \FF121332=-\FF121112=\FF212331=-\FF212111=\FF222002=-\FF222222 =\za,\\[2.4
mm]
  \FF111021=\FFb111201 =\zb,\qquad\
    \FF111001=-\FF111221 =\alpha \za,\\[2.4 mm]
  \FF112012=\FFb112232=\FFb211102=\FF211322 =\alpha \zb,\\[2.4 mm]
  \FF112032=-\FFb112212=\FFb211302=-\FF211122=\FFb122301=-\FF122121
     =\FF221031=-\FFb221211 =\alpha\omegae\za,\\[2.4 mm]
  \FFb121132=\FF121312 =\alpha\beta\bmegae \zb,\qquad\
    \FFb122101=\FF122321=\FF221011=\FFb221231 =\alpha\beta\omegaz \zb,\\[2.4
mm]
  \FF212311=\FFb212131 =\alpha\bmegae\omegaz \zb,\qquad
    \FF222022=\FFb222202 =\alpha\beta\omegae^2\omegaz^{} \zb,
   \\[-6 mm] {~}    \end{array}\labl{53}
The parameters in \erf{53} are restricted by
  \be  \alpha,\beta\in\{1,-1\}, \qquad |\omegaa|=|\omegae|=|\omegaz|=1, \ee
and
  \be  \ZA=\vert \zb\vert=\sqrt{\frac12(\sqrt5-1)}  \ee
(thus $\ZA$ is as in \erf{513}).
Similarly, the hexagon equations amount to 142 equations for the $F$s
and for 13 further variables $R$. The general solution is
  \be  \begin{array}{l}
  \RR110=\beta\zetae^3\zetA^2,\quad
  \RR112=\beta\zetae^3\zetA^{},\quad
  \RR121=\RR211=\RR222=\zetA, \quad
  \RR123=\RR213=\beta\zetae^2\zetA^2, \\[2.8 mm]
  \RR132=\RR312=\zetae,\quad
  \RR220=\zetA^2,\quad
  \RR231=\RR321=\beta\zetae^2,\quad
  \RR330=\beta\zetae^3.
    \\[-6 mm] {~}   \end{array}\labl{54}
The parameters $\zetae$ and $\zetA$ are phases with the specific values
  \be  \zetae^2=\alpha,\qquad
  \zetA={1\over 4}\cdot\left(1-\sqrt{5}\pm\ii\sqrt{10+2\sqrt{5}}\right). \ee
(thus $\zetA$ is the same as in \erf{514}).
Finally,
  \be  r=e_0^{}+\sqrt{(\sqrt5+1)/2}\,(\e111+\e211)+e_3^{}, \quad
       l=e_0^{}+\sqrt{(\sqrt5+1)/2}\,(\alpha\e111+\e211)+\alpha e_3^{}. \ee
Here and below we use the abbreviations
  \be  e_{rs}=e_r+e_s\,, \quad f_{rs}=e_r+\omegae\omegaz e_s \,, \ee
and
  \be  \quad \e{}ij=\e1ij+\e2ij\,, \quad f^{i,j}=\bmegae\e1ij+\omegae\e2ij \ee
for $i,j=1,2.$
The non-trivial statistics operators read as
  \small $$   \begin{array}{l}
  \stopt_{33}=\beta\zetae^3\cdot\UN, \\ {}\\[.4   mm]
  \stopt_{13}=\zetae\left( \begin{array}{cc}
    e_0+\alpha\bmegae\e111+\omegaa\omegae\e211+\alpha\bmegaa e_3^{}
    & \beta\bmegaz\e121\\[.9 mm]
    \beta\omegaz\e212
    & e_0^{}+\beta\bmegaz\e122+\beta\omegaz\e222+\alpha\bmegaa e_3^{}
    \end{array} \right),\\ {}\\[.4   mm]
  \stopt_{31}=\zetae\left(\begin{array}{cc}
    e_0^{}+\alpha\omegae\e111+\bmegaa\bmegae\e211+\alpha\omegaa e_3^{}
    & \beta\bmegaz\e221\\[.9 mm]
    \beta\omegaz\e112
    & e_0^{}+\beta\omegaz\e122+\beta\bmegaz\e222+\alpha\omegaa e_3^{}
    \end{array}\right),\\ {}\\[.4   mm]
  \stopt_{23}=\alpha\beta\left(\begin{array}{cc}
    e_0^{}+\omegaa\omegae\e111+\alpha\bmegae\e211+\alpha\bmegaa e_3^{}
    &0\\[.9 mm]
    0&e_0^{}+\beta\bmegae e_1^{}+\beta\omegae e_2+\alpha\bmegaa e_3^{}
    \end{array}\right),\\ {}\\[.4   mm]
  \stopt_{32}=\alpha\beta\left(\begin{array}{cc}
    e_0+\bmegaa\bmegae\e111+\alpha\omegae\e211+\alpha\omegaa e_3&0\\[.9 mm]
    0&e_0+\beta\omegae e_1^{}+\beta\bmegae e_2^{}+\alpha\omegaa e_3^{}
    \end{array}\right), \\ {}\\[.4 mm]
  \stopt_{11}=\beta\zetae^3\! \left(\begin{array}{cccc}
    \!\! \zetA^2 e_{03}-\zetA^3 \za e^{1,1}+\zetA \e222 \!\!
    &-\bmegae\zetA^4\zb\e212 &-\alpha\zetA^4\zb \e112 &0\\[.9 mm]
    -\zetA^4\Zb\omegae\e221
    &\! \zetA e_0-\zetA^3\za e^{11}-\za \e222 \! &0
    & \!\!\!\! -\zetA^4\zb(\alpha \e112+\bmegae\e212) \!\!\!\! \\[.9 mm]
    -\alpha\zetA^4\Zb \e121 &0 &\! \zetA(e_3^{}+\e111)-\za \e122 \!&0\\[.9 mm]
    0 &-\zetA^4\Zb(\alpha \e121+\omegae\e221)
    &0 &\zetA e_{03}-\za e^{22} \end{array}\right), \end{array}$$ 
  \be  \begin{array}{l}
  \stopt_{22}=\left(\begin{array}{cccc}
    \zetA^2 e_{03}-\zetA^3 \za e^{1,1}_{} &0 &0
    & \!\!\! -\beta\omegaz\zetA^4\zb(\bmegae\e111+\alpha\omegae^2\e211)
      \!\! \\[.9 mm]
    0 &\zetA e_{03}-\zetA^3\za e^{1,1}_{} &0
    & \!\!\! -\beta\omegaz\zetA^4\zb(\bmegae\e112+\alpha\omegae^2\e212)
      \!\! \\[.9 mm]
    0 &0 &\zetA e^{1,1}_{} &0\\[.9 mm]
    \!\! -\beta\bmegaz\zetA^4\Zb(\omegae\e111+\alpha\bmegae^2\e211)
    \! & \! -\beta\bmegaz\zetA^4\Zb(\omegae\e121+\alpha\bmegae^2\e221)
    \!\! &0 &\zetA e_{03}-\za e_{12} \end{array} \right),
    \\ {}\\[.4   mm]
  \stopt_{12}=\left(\begin{array}{cccc}
    \alpha\beta\zetA^2 e_{03}-\beta\zetA^3 \za f^{1,1}
    &-\alpha\beta\omegaz\zetA^4\zb\e212 &0 &-\beta\zetA^4\zb\e111\\[.9 mm]
    0 &-\beta\zetA^3\za f^{1,1}+\omegae\omegaz\zetA e_3 &0
    &-\beta\zetA^4\zb(\e112+\alpha\omegaz\e212)\\[.9 mm]
    \beta\bmegae\zetA\e212-\bmegaz\zetA^4\Zb\e221
    &\zetA e_0-\alpha\bmegae\za \e222 &\beta\bmegaz\zetA\e111 &0
    \\[.9 mm] -\alpha\zetA^4\Zb\e111
    &-\zetA^4\Zb(\alpha \e121+\bmegaz\e221) &0
    &\zetA f_{03}-\alpha \za(\omegae e_1+\bmegae\e222) \end{array}\right),
    \\ {}\\[.4   mm]
  \stopt_{21}=\left( \begin{array}{cccc}
    \!\! \alpha\beta\zetA^2 e_{03}-\beta\zetA^3 \za \bar f^{1,1} \!\! &0
    & \!\! \beta\omegae\zetA\e221-\omegaz\zetA^4\zb\e212 \!\!
    &-\alpha\zetA^4\zb\e111\\[.9 mm]
    -\alpha\beta\bmegaz\zetA^4 \Zb\e221
    & \!\! -\beta\zetA^3\za \bar f^{1,1}+\bmegae\bmegaz\zetA e_3 \!
    &\zetA e_0-\alpha\omegae\za\e222
    &-\zetA^4\zb(\alpha \e112+\omegaz\e212)\\[.9 mm]
    0 &0 &\beta\omegaz\zetA\e111 &0\\[.9 mm]
    -\beta\zetA^4 \Zb\e111
    &-\beta\zetA^4 \Zb(\e121+\alpha\bmegaz\e221) &0
    & \!\!\!\! \zetA f_{03}^*-\alpha \za(\bmegae e_1+\omegae\e222) \!\!\!
    \end{array}\right). \\[-4 mm] {}\\{}
  \end{array} \normalsize \ee  \normalsize 
The left inverses are $\Phi_0(f)=\id_H$,
  \be  \Phi_3(f)=\left\{ \begin{array}{ll} e_3, & f=e_0,\\[.9 mm]
  \e2ij, & f=\e1ij,\\[.9 mm] \e1ij, &  f=\e2ij,\\[.6 mm]
  e_0, &  f=e_3, \end{array} \right.  \ee
and
  \be  \begin{array}{l}
  \Phi_1([f])=\left\{ \begin{array}{ll}
   \Za\cdot \e1kl, &    [f]={[e_0]}^{k,l},\\[.7 mm]
   \za\cdot \e2ij, &   [f]={[\e1ij]}^{2,2},\\[.7 mm]
   e_0, & [f]={[\e111]}^{1,1},\\[.7 mm] e_3, & [f]={[\e211]}^{1,1},\\[.7 mm]
   \za\cdot \e1kl, &    [f]={[\e222]}^{k,l},\\[.7 mm]
   \Za\cdot \e2kl, &    [f]={[e_3]}^{k,l},\\[.7 mm]
   0, &    {\rm otherwise,} \end{array} \right. \qquad
  \Phi_2([f])=\left\{ \begin{array}{ll}
   \Za\cdot \e2kl, &    [f]={[e_0]}^{k,l},\\[.7 mm]
   \za\cdot \e1ij, &   [f]={[\e1ij]}^{2,2},\\[.7 mm]
   e_3, & [f]={[\e111]}^{1,1},\\[.7 mm] e_0, & [f]={[\e211]}^{1,1},\\[.7 mm]
   \za\cdot \e2ij, &    [f]={[\e2ij]}^{2,2},\\[.7 mm]
   \Za\cdot \e1kl, &    [f]={[e_3]}^{k,l},\\[.7 mm]
   0, &    {\rm otherwise.} \end{array} \right.
   \end{array}\ee
For the statistics parameters we then find
  \be  \lambda_0=\UN,\qquad \lambda_1=-\beta\zetae^3\zetA^3\za\cdot\UN,\qquad
  \lambda_2=-\zetA^3\za\cdot\UN,\qquad \lambda_3=\beta\zetae^3\cdot\UN.
  \labl{615}
Thus the statistical dimensions are
  \be  d_0=1,\qquad d_1={1+\sqrt{5}\over 2},\qquad
  d_2={1+\sqrt{5}\over 2},\qquad   d_3=1, \ee
so that $\sum_r d_r^2=5+\sqrt5$.
Finally, the monodromy matrix reads
  \be  y= \left( \begin{array}{cccc} 1 & \sqf & \sqf & 1 \\[.9 mm]
  \sqf & -\alpha & -1 & \alpha\,\sqf\\[.9 mm] \sqf & -1 & -1 & \sqf\\[.9 mm]
  1 & \alpha\,\sqf & \sqf & \alpha \end{array}\right) , \ee
which is non-degenerate iff $\alpha=-1$.\\
According to \erf{615} we are  left with two parameters $\beta\zetae^3$
and $\zetA^3$, so that we  arrive at the following candidates for \rha s:
  \be \begin{tabular}{rc|cccccc|r} \multicolumn{1}{c}{$\beta\zetae^3\!\!\!$}
  & \multicolumn{1}{c|}{$-\zetA^3$}
  & $w_0$ & $w_1$ & $w_2$ & $w_3$ & $|\sigmA|^2$ & $c$
  & \interpr \\[.7 mm] \hline &&&&&&&\\[-2.7 mm]
  \ii & $\exp(2\pi\ii\frac35)$ & 0 & $\frac{17}{20}$ & $\frac35$ & $\frac14$
                & $  5+\sqrt5 $ & $\frac{31}5$& \hzz A11F41 \\[1.3 mm]
  $-1$& $\exp(2\pi\ii\frac35)$ & 0 & $\frac1{10}$  &   $\frac35$ & $\frac12$
                & 0             & .           & \degen      \\[1.3 mm]
$-\ii$& $\exp(2\pi\ii\frac35)$ & 0 & $\frac7{20}$  &   $\frac35$ & $\frac34$
                & $  5+\sqrt5 $ & $\frac{21}5$& \hzz E71F41,\ \hz C31 \\[1.3
mm]
  1   & $\exp(2\pi\ii\frac35)$ & 0 & $\frac35$     &   $\frac35$ & 0
                & $2(5+\sqrt5)$ & .           & \degen      \\[1.3 mm]
  \ii & $\exp(2\pi\ii\frac25)$ & 0 & $\frac{13}{20}$ & $\frac25$ & $\frac14$
                & $  5+\sqrt5 $ & $\frac{19}5$& \hzz A11G21 \\[1.3 mm]
  $-1$& $\exp(2\pi\ii\frac25)$ & 0 & $\frac9{10}$  &   $\frac25$ & $\frac12$
                & 0             & .           & \degen      \\[1.3 mm]
$-\ii$& $\exp(2\pi\ii\frac25)$ & 0 & $\frac3{20}$  &   $\frac25$ & $\frac34$
                & $  5+\sqrt5 $ & $\frac95$   & \hzz E71G21,\ \hz A13 \\[1.3
mm]
  1   & $\exp(2\pi\ii\frac25)$ & 0 & $\frac25$     &   $\frac25$ & 0
                & $2(5+\sqrt5)$ & .           & \degen
  \end{tabular} \ee
Thus we found four inequivalent \rha s (all of which have $\alpha=\zetae
^2=-1$) as well as four degenerate algebras (with $\alpha=1$).
All these solutions are precisely the tensor products of the
solutions that we found previously for the $\zet_2$ and \LY fusion
rules. In contrast, in addition to the tensor products of the \cft\ models
that have the $\zet_2$ and \LY \furu s, there exist also \cfts\ with
the tensor product \furu s that are themselves not tensor products of
simpler \cfts, namely the \wz A13 and \wz C31 \wzwts.

\sect{Outlook} \label{secnp}

A major goal in the \DHR programme for the reconstruction of \qfts\
is to determine the internal symmetry from the observable algebra
\OA\ and the category of its physical \rep s (or, equivalently, the
category of localized transportable automorphisms $\rho$ of \OA).
As already remarked in the introduction, the completion of this
programme in $D\le2$ dimensions is still an open problem.
Our analysis in section 3 of the structure of \rha s in practice
trivializes this reconstruction problem in the case where \OA\
possesses only a finite number of superselection sectors.
Indeed, the essential data characterizing the category of physical
\rep s of \OA\ consists of the \furu s \npqr\ and the components
\Fpqr\ and $\R pqt\alpha\beta$ of the fusion and braiding matrices.
As follows from the results of section 3 (compare also \cite{SV}),
these data allow to reconstruct
a \rha\ with the same category of \rep s, and this \rha\ is unique
up to a gauge choice and up to the choice of the dimensionalities $n_r$.

The main task that has still to be performed is then the difficult
problem of classifying \rha s.
One of the results of the present paper is the complete classification of
\rha s with a small number of sectors, which was presented in sections
5 and 6.  In the same spirit, a classification of all \rha s could proceed
as follows. The first step is the classification of (modular) fusion
rings. So far only the fusion rings with three or less generators
have been classified, while the general case is far from
being complete \cite{FRke,F}. (Assuming that to any modular fusion ring
there exists at least one rational Hopf algebra whose character ring
coincides with the fusion ring, one can in fact try to employ the methods
presented here to gain new information about the classification of
fusion rings.) The second step is the one considered on several examples
above: given a fusion ring $\hh$, find all rational Hopf algebras \H\
(modulo the gauge freedom) that have $\hh$ as their character ring.
We have approached this problem rather directly, namely by solving the
polynomial equations \erf{penf} and \erf{hexf}. Although with the
computer programs mentioned in
section 6 this approach will still work for larger fusion rings than the
ones considered here, this is in any case reasonable only for small
enough $|\hh|$. However, the rational Hopf algebras
having the same character ring $\hh$ differ essentially only by the values
of the statistics phases $\omega_r$, $r\in\hh$. Therefore it is
plausible that one can bypass the difficult problem of solving
the polynomial equations and find methods producing directly
the possible statistical phases. A step in this direction would
be the understanding of the pentagon equation as a 3-cocycle condition
in an appropriate non-abelian cohomology. In the case where
there are only abelian sectors, the fusion ring is in fact the group
ring of an abelian group, and the cohomological interpretation
of the pentagon identity is already known \cite{mose}.
Also note that the five factors in the pentagon identity can be
associated to the five three-dimensional basic simplices (i.e., tetrahedra)
in the boundary of the four-dimensional basic simplex, so that
the pentagon identity expresses the `contractibility'
of the four-dimensional simplex.

Among the rational field theories, there are in particular the rational
chiral \cfts. Accordingly, from the results reported in section 2 it
follows that the classification of rational Hopf algebras leads to a partial
classification of unitary rational \cfts. Thus a natural question is how
well can rational Hopf algebras distinguish between rational field theories,
or in particular between rational chiral conformal field theories.
A rational Hopf algebra encodes the information about
the fusion rules, the modular group representation, the statistical phases,
and the braid group representation of a model. However, the central
charge \erf c is determined only modulo 8; similarly,
via the spin-statistics theorem, the statistical phases
$\omega_r$ determine only the fractional part of the spins, respectively
of the conformal weights in chiral \cft.
Indeed, there always exist infinitely many different theories sharing a
given rational Hopf algebra. For instance, according to \erf{iscl}
the $(B_{n+8m})_1$ WZW theories with fixed $n$ and different values
of $m$ all correspond to the same \rha. Further, by tensoring any \cft\
model by the $(E_8)_1$ WZW theory, one obtains a theory with the
same \rha\ as the original model.
 \futnote{Note that \cft\ models with
different chiral symmetry algebras are not automatically
distinct conformal field theories. This is because the relevant object is the
chiral symmetry algebra modulo the so-called annihilating ideal, i.e.\
the ideal consisting of all elements that are represented by
zero in each physical representation \cite{annih}. But in the present
situation the relevant models have different central charges $c$
and hence are certainly distinct theories.}
 As a \rha, being the global symmetry algebra, reflects only the structure
of the field algebra modulo the
chiral symmetry algebra, i.e.\ the superselection sectors, it cannot make
a distinction between different chiral symmetry algebras whose \rep s
lead to equivalent braided monoidal rigid $C^*$-categories.

\vskip 2mm
Another issue analyzed in this paper is the relevance of the
dimensionalities $n_p$ of the simple summands $M_{n_p}$ of \h. We
have seen that all important structural information on \rha s
can be formulated independently of the actual values of these numbers.
Also, as already pointed out in the introduction, in order to produce
non-integral statistical dimensions $d_r$
one must use a unit non-preserving coproduct
 \futnote{There are lots of examples of \RHA s with coassociative and unit
preserving coproduct. Namely, every double ${\cal D}(G)$ of a finite
group $G$ is a \RHA. These algebras describe the global symmetries of
$G$-orbifold models \cite{orbif} and $G$-spin chains \cite{SV'}.}
 in $H$. It is actually quite remarkable that one can arrive at
non-integral statistical dimensions and even non-integral indices
$d_r^2$ by employing only finite-dimensional constructions. However,
the corresponding dimensionalities $n_r$ of the simple
ideals of $H$ are not fixed by the properties of rational Hopf algebras.
The only requirements are $n_0=1,\, n_r=n_{\bar r}$ and the inequalities
  \be  n_pn_q\geq\sum_{r\in\hh} \npqr\,n_r \qquad {\rm for\ all}\
  p,q\in\hh \,. \labl{81}
For example, we have checked that
for arbitrary integer $n\geq 2$, the algebras $H_n:=M_1\oplus M_n$
can be endowed with a \RHA\ structure that possesses the \LY fusion
rules and reproduces the same statistical parameters and
monodromy matrix as obtained in \erf{517} and \erf{518}. Of course, this
is a direct consequence of the general statement (see subsection 4.1) that the
coassociator, the cocommutator and the intertwiners $l$ and $r$
together with the triangle, square, pentagon and hexagon
identities (we are working in a $\lambda=\rho=\UN$ gauge) can be
rewritten in an $n_r$-independent form.

A natural guess for a general prescription for the minimal allowed
choice of the $n_r$ is
  \be  d_r\leq n_r< d_r+1,\qquad r\in\hh \,. \labl{82}
This minimal choice is possible in all cases that we discussed in
sections 5 and 6. Moreover, in the case of integral statistical
dimensions the minimal choice \erf{82} is allowed as well, as follows
from the general relations
  \be  d_pd_q=\sum_{r\in\hat H} N_{pq}^{\ \,r}d_r,\qquad p,q\in\hh, \ee
and from the fact that unit preserving coproducts (or,
equivalently, unit preserving amplimorphisms of a
\findim\ algebra \cite{SV'}) lead to an index $d_r^2$ which is
equal to the dimension $n_r^2$ of the amplification space.
However, unfortunately we were not able to prove that the (would-be
unique) choice \erf{82} always provides a solution to \erf{81}.

Of course this issue is strongly connected to the field algebra
that one would like to reconstruct from observable data.
In space-time dimensions $D>2$ \cite{DR}, the integral statistical
dimension $d_r$ plays simultaneously the roles of the dimension
of the relevant representation of the global symmetry group and of
the multiplicity of the corresponding sector of the
observables in the representation of the field algebra, that is
$d_r=n_r$. Obviously, this can no longer be the case if we have
non-integral statistical dimensions.

Maybe the requirement of some kind of weak irreducibility of the
$\vert\hh\vert$-coloured braid representation over $H$ that is
provided by the statistics operators, can restrict the
multiplicities in this case. By {\em weak irreducibility\/}
of $B_n(\vert\hh\vert)$ we mean that there does not exist any
proper semisimple subalgebra $G$ of $H$ with $|\hat G|=|\hh|$ such that
  \be  (\UN_G\otim \one_s\otim \one_r)\cdot\stopt_{rs}
  \cdot(\UN_G\otim \one_r\otim \one_s)=\stopt_{rs} \ee
(with $\one_r$ the unit matrix in $M_{n_r}(\complex)$)
is valid for some $r,s\in\hh$. At least in the case of the
mentioned examples of $H_n=M_1\oplus M_n$ such a subalgebra $G$ exists
for $\stopt_{11}$; namely, $G=M_1\oplus M_2$ for all choices $n>2$,
and therefore the requirement of weak irreducibility
over \H\ imposes the choice of $H_2$. The weak irreducibility
is automatically fulfilled for any \RHA\ with a coassociative and unit
preserving coproduct, since in those cases $\stopt_{rs}$ is a
scalar matrix multiple of the unit element of $H$ for all $r,s\in\hh$.

Finally, we would like to note that the only known field theoretical models
with non-integral statistical dimensions are chiral conformal
field theories. Therefore it might happen that non-integral
statistical dimensions can only occur in space-times with
non-trivial topology (such as the circle $S^1$ in the case of chiral
\cft), which does not admit a simple universal observable algebra
and proper asymptotic particle states.

   \def\tmis {{\bf TITLE MISSING}}
   \newcommand{\wb}{\,\linebreak[0]} \def\wB {$\,$\wb}
   \newcommand{\J}[1]     {{{#1}}\vyp}
   \newcommand{\Jj}[1]    {{{#1}}\vyP}
   \newcommand{\JJ}[1]    {{{#1}}\vyp}
   \newcommand{\PRep}[2]  {preprint {#1}}
   \newcommand{\PhD}[2]   {Ph.D.\ thesis (#1)}
   \newcommand{\Erra}[3]  {\,[{\em ibid.}\ {#1} ({#2}) {#3}, {\em Erratum}]}
   \newcommand{\BOOK}[4]  {{\em #1\/} ({#2}, {#3} {#4})}
   \newcommand{\inBO}[7]  {in:\ {\em #1}, {#2}\ ({#3}, {#4} {#5}), p.\ {#6}}
   \newcommand{\inBOnoeds}[6]  {in:\ {\em #1} ({#2}, {#3} {#4}), p.\ {#5}}
   \newcommand{\vyp}[4]   {\ {#1} ({#2}) {#3}}
   \newcommand{\vyP}[3]   {\ ({#1}) {#2}}
   \newcommand{\vypf}[5]  {\ {#1} [FS{#2}] ({#3}) {#4}}

   \renewcommand{\J}[5]   {{\sl #5}, {#1} {#2} ({#3}) {#4}}
   \renewcommand{\JJ}[5]  {{\sl #5}  {#1} {#2} ({#3}) {#4}}
   \renewcommand{\PRep}[2]{{\sl #2}, preprint {#1}}
   \renewcommand{\inBO}[7]{{\sl #7}, in:\ {\em #1}, {#2}\ ({#3}, {#4} {#5}),
                          p.\ {#6}}

   \def\acam  {Acta\wB Appl.\wb Math.}
   \def\acma  {Acta\wB Math.}
   \def\adma  {Adv.\wb Math.}
   \def\anma  {Ann.\wb Math.}
   \def\anop  {Ann.\wb Phys.}
   \def\asen  {Ann.\wb Sci.\wb Ec.\wb Norm.\wb Sup\'er.}
   \def\bams  {Bull.\wb Amer.\wb Math.\wb Soc.}
   \def\blms  {Bull.\wB London\wB Math.\wb Soc.}
   \def\bsbm  {Bol.\wb Soc.\wb Bras.\wb Math.}
   \def\bsmf  {Bull.\wb Soc.\wb Math.\wB de\wB France}
   \def\busm  {Bull.\wb Sci.\wb Math.}
   \def\coia  {Com\-mun.\wB in\wB Algebra}
   \def\coma  {Con\-temp.\wb Math.}
   \def\comp  {Com\-mun.\wb Math.\wb Phys.}
   \def\cpma  {Com\-pos.\wb Math.}
   \def\crap  {C.\wb R.\wb Acad.\wb Sci.\wB Paris}
   \def\foph  {Fortschr.\wb Phys.}
   \def\fuaa  {Funct.\wb Anal.\wb Appl.}
   \def\ijmb  {Int.\wb J.\wb Mod.\wb Phys.\ B}
   \def\ijmc  {Int.\wb J.\wb Mod.\wb Phys.\ C}
   \def\ijmp  {Int.\wb J.\wb Mod.\wb Phys.\ A}
   \newcommand{\ilag}[2] {\inBO{Infinite-\dimn al Lie \A s and Groups {\rm
              [Adv.\ Series in Math.\ Phys.\ 7]}} {V.G.\ Kac, ed.} \WS\Si{1989}
              {{#1}}{{#2}}}
   \def\inma  {Invent.\wb math.}
   \def\jams  {J.\wb Amer.\wb Math.\wb Soc.}
   \def\jgap  {J.\wb Geom.\wB and\wB Phys.}
   \def\joal  {J.\wB Al\-ge\-bra}
   \def\jodg  {J.\wb Diff.\wb Geom.}
   \def\jofa  {J.\wb Funct.\wb Anal.}
   \def\jopa  {J.\wb Phys.\ A}
   \def\jomp  {J.\wb Math.\wb Phys.}
   \def\lemp  {Lett.\wb Math.\wb Phys.}
   \def\lenc  {Lett.\wB Nuovo\wB Cim.}
   \def\leni  {Lenin\-grad\wB Math.\wb J.}
   \def\maan  {Math.\wb Annal.}
   \def\mams  {Memoirs\wB Amer.\wb Math.\wb Soc.}
   \newcommand{\mapx}[2] {\inBO{Mathematical Physics X}
              {K.\ Schm\"udgen, ed.} \SV\Be{1992} {{#1}}{{#2}} }
   \def\mpla  {Mod.\wb Phys.\wb Lett.\ A}
   \newcommand{\mqft}[2] {\inBO{Modern Quantum Field Theory}
              {S.\ Das, A.\ Dhar, S.\ Mukhi, A.\ Raina, and A.\ Sen, eds.}
              \WS\Si{1991} {{#1}}{{#2}}}
   \def\npbf  {Nucl.\wb Phys.\ B\vypf}
   \def\npbp  {Nucl.\wb Phys.\ B (Proc.\wb Suppl.)}
   \newcommand{\nspq}[2] {\inBO{
              New Symmetry Principles in Quantum Field Theory}
              {J.\ Fr\"ohlich et al., eds.} \PL\NY{1992} {{#1}}{{#2}}}
   \def\nupb  {Nucl.\wb Phys.\ B}
   \def\paaa  {Proc.\wb Amer.\wb Acad.\wB Arts\wB Sci.}
   \def\pams  {Proc.\wb Amer.\wb Math.\wb Soc.}
   \def\pcps  {Proc.\wB Cam\-bridge\wB Philos.\wb Soc.}
   \def\phlb  {Phys.\wb Lett.\ B}
   \def\pkna  {Proc.\wb Kon.\wb Ned.\wb Akad.\wb Wetensch.}
   \def\plms  {Proc.\wB Lon\-don\wB Math.\wb Soc.}
   \def\pnas  {Proc.\wb Natl.\wb Acad.\wb Sci.\wb USA}
   \def\prsa  {Proc.\wb Roy.\wb Soc.\wB Ser.$\,$A}
   \def\prtp  {Progr.\wb Theor.\wb Phys.}
   \def\pspm  {Proc.\wb Symp.\wB Pure\wB Math.}
   \def\ptps  {Progr.\wb Theor.\wb Phys.\wb Suppl.}
   \def\ptrs  {Phil.\wb Trans.\wb Roy.\wb Soc.\wB Lon\-don}
   \def\remp  {Rev.\wb Mod.\wb Phys.}
   \def\rpmp  {Rep.\wb Math.\wb Phys.}
   \def\rmap  {Rev.\wb Math.\wb Phys.}
   \def\slnm  {Sprin\-ger Lecture Notes in Mathematics}
   \def\slnp  {Sprin\-ger Lecture Notes in Physics}
   \newcommand{\Suse} [2] {\inBO{The Algebraic Theory of Superselection
              Sectors.\ Introduction and Recent Results} {D. Kastler,
              ed.} \WS\Si{1990} {{#1}}{{#2}} }
   \def\tams  {Trans.\wb Amer.\wb Math.\wb Soc.}
   \def\thmp  {Theor.\wb Math.\wb Phys.}
   \def\topo  {Topology}
   \def\AMS    {{American Mathematical Society}}
   \def\AP     {{Academic Press}}
   \def\AW     {{Addi\-son\hy Wes\-ley}}
   \def\BC     {{Ben\-jamin\,/\,Cum\-mings}}
   \def\BIR    {{Birk\-h\"au\-ser}}
   \def\CUP    {{Cambridge University Press}}
   \def\CUPC   {{Cambridge University Press}}
   \def\DP     {{Dover Publications}}
   \def\GB     {{Gordon and Breach}}
   \def\JW     {{John Wiley}}
   \def\KLU    {{Kluwer Academic Publishers}}
   \def\MD     {{Marcel Dekker}}
   \def\MGH    {{McGraw\,\hy\,Hill}}
   \def\NH     {{North Holland Publishing Company}}
   \def\OUP    {{Oxford University Press}}
   \def\PL     {{Plenum}}
   \def\PUP    {{Princeton University Press}}
   \def\SV     {{Sprin\-ger Verlag}}
   \def\WI     {{Wiley Interscience}}
   \def\WS     {{World Scientific}}
   \def\Be     {{Berlin}}
   \def\Ca     {{Cambridge}}
   \def\NY     {{New York}}
   \def\pR     {{Princeton}}
   \def\Si     {{Singapore}}
\def\A       {Algebra}
\def\aff     {affine Lie algebra}
\def\alg     {algebra}
\def\Class   {Classification\ }
\def\class   {classification}
\def\Con     {Conformal\ }
\def\con     {conformal\ }
\def\cua     {current algebra}
\def\dimn    {dimension}
\def\emt     {energy-momentum tensor}
\def\enva    {enveloping algebra}
\def\eq      {equa\-tion}
\def\fts     {field theories}
\def\hopf    {Hopf algebra}
\def\ide     {identification}
\def\jf      {J.\ Fuchs}
\def\km      {Kac\hy Moody}
\def\kze     {Knizh\-nik\hy Za\-mo\-lod\-chi\-kov equation}
\def\lie     {Lie algebra}
\def\q       {quantum\ }
\def\Q       {Quantum\ }
\def\stc     {statistic}
\def\sym     {symmetry}

\vskip 2cm \small
\noindent{\bf Acknowledgement.}\\[.5 mm] We thank the universities of
Heidelberg and Mannheim for access to their computer facilities.

\typeversion
\end{document}